\documentclass[11pt,reqno]{article}
\usepackage{graphicx,mathtools,amsmath,amsthm,amssymb,amsfonts,fullpage} % Required for inserting images
\usepackage{hyperref}
\usepackage{bm}
\usepackage{enumitem}
\usepackage{authblk}
\usepackage{xcolor}
\newtheorem{theorem}{Theorem}

\newtheorem{proposition}[theorem]{Proposition}
\newtheorem{lemma}[theorem]{Lemma}

\newtheorem{corollary}[theorem]{Corollary}

\newtheorem{fact}[theorem]{Fact}

\newtheorem*{claim*}{Claim}
\theoremstyle{remark}
\newtheorem*{remark*}{Remark}
\newtheorem{remark}[theorem]{Remark}

\DeclareMathOperator{\diag}{diag}
\DeclareMathOperator{\Comp}{Comp}

\usepackage{thmtools,thm-restate}
\mathtoolsset{showonlyrefs}

\usepackage{tikz}
\usetikzlibrary{arrows.meta,decorations.markings}
\definecolor{flow}{RGB}{39,112,153}
\definecolor{rim}{RGB}{38,65,86}

\newcommand{\R}{\mathbb{R}}
\newcommand{\E}{\mathbb{E}}
\newcommand{\Cov}{\operatorname{Cov}}
\newcommand{\diver}{\operatorname{div}}

\newcommand{\norm}[1]{\left\lVert #1\right\rVert}

\newcommand{\bS}{\mathbb S}

\newcommand{\op}{\mathrm{op}}
\DeclareMathOperator{\Var}{Var}

\DeclareMathOperator{\Lip}{Lip}

\newcommand{\one}{\mathbf 1}

\title{Beyond the Bochner Method: Optimal Spectral Gaps for Block Dynamics of Spin Systems like the $O(N)$ Model}
\author[1]{Jonathan Leake}
\author[2]{Anqi Li}
\affil[1]{University of Waterloo, \textsf{jonathan.leake@uwaterloo.ca}}
\affil[2]{University of Washington, \textsf{aqli@cs.washington.edu}}
\date{}

\begin{document}

\maketitle

\begin{abstract}
    Let $\mu$ be a probability distribution on a spin system with $n$ sites and possibly infinite spin spaces. We prove local-to-global spectral-gap bounds for block Glauber dynamics.
Given a symmetric, entrywise nonnegative matrix $A$ with zero diagonal
and $\lambda_{\max}(A)<1$ encoding \emph{local} spectral bounds for a
spin system, we derive \emph{global} spectral-gap bounds for block
dynamics on its marginal distributions. These bounds are sharp:
the Gaussian system with covariance $(I-A)^{-1}$ attains equality
for every choice of disjoint blocks. Our proof extends the Bochner method through an abstract inequality for orthogonal projections on Hilbert spaces.

    As an application, we prove that Glauber dynamics for the $O(N)$ model, with $N\geq 2$, arbitrary external fields, and couplings of either sign, has spectral gap at least $(1-\|J_{\mathrm{abs}}\|_{\mathrm{op}}/N)/n$, where $J$ is the interaction matrix and $J_{\mathrm{abs}}=(|J_{ij}|)$. 
    
    We also obtain spectral bounds for swap walks and local spectral expansion of partite simplicial complexes. For complexes formed from disjoint groups of ranks in random flags of subspaces over $\mathbb F_q$, we obtain a one-sided local spectral expansion bound of $2\sqrt q/(q+1)$. This bound
    is independent of the number and sizes of each group, and is nontrivial for every prime power $q$.

    Finally, for random forests $F$ of any finite graph with arbitrary positive edge weights, we prove that
    \[
    \Pr(e,f\in F)\le \frac{4}{3} \Pr(e\in F)\Pr(f\in F)
    \]
    for distinct edges $e,f$, improving the factor 2 supplied by general log-concavity bounds for matroid independent sets. The argument combines our extended Bochner method with the analysis of an auxiliary real stable polynomial.
\end{abstract}

\section{Introduction}

Spectral gaps connect expansion, functional inequalities, and the convergence of random walks. On graphs, they quantify connectivity and control the rate at which a walk approaches equilibrium. Through Poincar\'e inequalities, they also bound the variance of functions in terms of their local variation, providing a connection to correlations and fluctuations. These perspectives have made spectral estimates useful throughout combinatorics, probability, and statistical physics, with applications to sampling and approximate counting, the dynamics of spin systems, and high-dimensional expansion. A recurring difficulty is that the relevant state spaces can be enormous or even continuous, even when their defining interactions are simple. This motivates a local-to-global approach: obtain spectral information about a complicated system by analyzing smaller, more tractable pieces.

Trickle-down theorems in high-dimensional expansion provide one realization of this local-to-global approach (see \cite{Opp18} and its extensions \cite{ALO21,AKV24,LLO25}). The local structure of a simplicial complex is described by its links, which record the ways a fixed face can be extended to larger faces. When a finite spin system is encoded as a simplicial complex, passing to a link corresponds to conditioning on the spins at specified sites. Under suitable connectivity assumptions, spectral expansion of low-dimensional links propagates to higher-dimensional links, leading to spectral bounds for random walks on the complex.

\paragraph{The Bochner method.}

In very recent times, many spectral-gap arguments like those utilized in trickle-down theorems now admit substantially simpler proofs via the Bochner method. The technique is similar, local expansion implies global expansion, but the main difference is in the simplicity and uniformity of its proof. This method has been utilized in a huge number of new results (see for instance \cite{GJMPPS26, GZ26, Wan26, CL26c, CL26b, BLO26, Haq26, LO26} and also several recent papers by OpenAI). We note that the new explosion of results is mainly driven by AI, and even the general power of the technique itself within the context of trickle-down theorems was not fully realized until it was suggested by AI.

The Bochner method is not a new technique (it is called ``Bochner'' after all). The name derives from \cite{Boc46,Lic58}, where a similar analysis is used to prove spectral gaps for the Laplacian of a Riemannian manifold (see also \cite{BGL14}). Various papers since then have made connections to graphs, expansion, spectral gaps, and beyond \cite{BCDP06,BGL14}. The general idea is to represent an eigenvalue gap by a matrix inequality, for example $L^2 - \lambda L \succeq 0$ for the Laplacian $L$. Such an inequality equivalently says that $L$ has no eigenvalues in $(0,\lambda)$, and connectivity of the graph implies the $0$ eigenvalue is simple. By representing $L^2 - \lambda L$ as $\sum_i L_i^2 - \lambda_i L_i$ using certain local Laplacians $L_i$, one immediately derives a global spectral gap from local ones. We direct the reader to the very short paper \cite{CL26a} for further details.

\paragraph{Beyond the Bochner method.}

In the context of spin systems (which we define below), the usual Bochner argument yields a spectral gap for the whole system from conditional estimates on pairs of sites. In this paper, we extend this method to handle a much wider range of conclusions, allowing for arbitrary marginalization and groupings of sites, all following from the same local information. Moreover, our new spectral gap bounds are tight: for each admissible matrix of local bounds, a single Gaussian system attains all of these bounds simultaneously.

Our applications include spectral gap bounds and sampling algorithms for the $O(N)$ model (Section~\ref{sec:main-ON}), applications to swap walks and local spectral expansion (Section~\ref{sec:intro-hdx}), and improved negative correlation bounds for pairs of edges of uniformly random forests in a graph (Section~\ref{sec:main-neg-corr}).

To achieve our results, we abstract the content of the Bochner method. We prove a purely linear algebraic statement, which when applied to local Laplacians of a given graph, immediately recovers the spectral gap conclusions of both the Bochner method and our new results. Our hope for this formulation is that we are getting closer to some core underlying mathematical framework that might bring together the various local-to-global results spread across the math and TCS literature. This is an important goal in itself in the AI age, where specific theorems may become cheaper than the techniques used to conceptually simplify and unify them.

We now discuss these results and applications in more detail. We first describe the setup in terms of spin systems, and then give our main results and applications. After that, we will describe the purely linear algebraic formulation of our technical results, and how it yields our applications.

\subsection{Spin systems and Glauber dynamics}

A \textbf{spin system} is a graph $G=(V,E)$ with \textbf{sites} $V$ and \textbf{spins} $S_v$ associated to each site $v \in V$. A \textbf{configuration} of the spin system is then a valid choice of spin $s_v \in S_v$ for every $v \in V$. The validity of the choices of spins are determined by certain rules imposed along the edges of the graph.

We describe two basic examples. Given any graph, let $S_v$ be a set of colors for all $v \in V$, and let the edges impose the rule that no adjacent vertices are allowed to have the same color (spin). Then the set of configurations is the set of proper colorings of the graph. As another example, let $S_v = \{0,1\}$ for all $v \in V$, and let the edges impose the rule that adjacent vertices cannot both be equal to $1$. Then the set of configurations is the set of (indicator functions of) independent sets of the graph. Both models are classical settings for sampling via Glauber dynamics \cite{J95,LV99}. Other important examples of spin systems (like the $O(N)$ model we study) allow for infinite sets of spins.

A spin system is also equipped with a measure $\mu$ on the set of all its configurations. In this paper we study the \textbf{Glauber dynamics} of a given spin system, which is a natural random walk on its configurations. This is how it works: Given a configuration, choose a uniformly random vertex and forget its spin. Then, sample a new spin for that vertex according to the $\mu$-conditional distribution. It is a standard fact that this process converges to the distribution $\mu$.

Let $P_{\mathrm{Glaub}}$ be the probability transition matrix of the Glauber random walk, so that for configurations $\sigma,\tau$, the entry $P_{\mathrm{Glaub}}(\sigma,\tau)$ is the probability that the walk transitions from $\sigma$ to $\tau$ in one step. The \textbf{mixing time} of Glauber dynamics is defined to be
$$\inf\Big\{
t \in \mathbb Z_{\geq 0} :
\sup_{\sigma \in \mathcal S}
\left| P_{\mathrm{Glaub}}^t(\sigma,\cdot)-\mu \right|_{\mathrm{TV}}
\leq \frac14
\Big\}.$$ When the set of spins $S_v$ is finite for all $v \in V$, the spectral gap of the Glauber dynamics is defined as expected as a literal gap between eigenvalues. The rows of this matrix clearly sum to $1$, and thus the maximum eigenvalue is equal to $1$. The spectral gap is then equal to $\lambda_2(I - P_{\mathrm{Glaub}})$.
    
However, when considering infinite spaces of spins (e.g. when $S_v$ is the unit sphere of $\R^m$), this interpretation breaks down. In this case one needs to consider operators on infinite-dimensional Hilbert spaces, and not all points of the spectrum of an operator are eigenvalues. This can be remedied, by instead defining the spectral gap via quadratic inequality discussed above, now for the Laplacian (under a certain connectedness assumption). That is, for the Laplacian $L_{\mathrm{Glaub}} = I - P_{\mathrm{Glaub}}$, the \textbf{spectral gap} is defined to be the maximal $\lambda \geq 0$ such that $L_{\mathrm{Glaub}}^2 - \lambda L_{\mathrm{Glaub}} \succeq 0$, where we set $\lambda = 0$ if $\dim(\ker(L_{\mathrm{Glaub}})) \geq 2$ (the Laplacian condition for disconnectedness). We refer to this as the spectral gap of $L_{\mathrm{Glaub}}$ (or $P_{\mathrm{Glaub}}$), and write $\lambda_2(L_{\mathrm{Glaub}}) = \lambda$. In the finite spins case, this is equivalent to the previous definition.

\paragraph{Using the Bochner method.}

The Bochner method then gives a simple method for proving spectral gaps in the case of general spin systems (with not necessarily finite sets of spins). In this section we give the main result in this direction and refer the reader to \cite{GZ26} for the original statement, or \cite{CL26a} for the clean and direct proof of this fact.

First we need a bit of extra notation. A \textbf{pinning} of a spin system is a fixing of some of the spins. Note that a pinning is a spin system on the unpinned vertices, with distribution $\mu$ conditioned on the pinned spins. Thus, we can still consider the Glauber dynamics and define $P_{\mathrm{Glaub}}^\tau$ and $L_{\mathrm{Glaub}}^\tau$ for a given pinning $\tau$.

\begin{theorem}[\cite{GZ26,CL26a}, see also \cite{LO25}] \label{thm:main-prev-Bochner}
    Fix a spin system on $G=(V,E)$ with configurations distributed according to $\mu$, such that $\ker(L_{\mathrm{Glaub}})$ consists only of constant functions (connectedness). Fix also a $V \times V$ symmetric matrix $A$ with non-negative entries and zero diagonal such that $\lambda_{\max}(A) < 1$. If for all $u \neq v$ in $V$ and all pinnings $\tau$ of $V \setminus \{u,v\}$ we have
    \[
        \lambda_2(L_{\mathrm{Glaub}}^\tau) \geq \frac{1-a_{uv}}{2},
    \]
    then
    \[
        \lambda_2(L_{\mathrm{Glaub}}) \geq \frac{1-\lambda_{\max}(A)}{|V|}.
    \]
\end{theorem}

Note that, under the appropriate connectedness assumption, the conclusion of Theorem~\ref{thm:main-prev-Bochner} extends to any intermediate pinning by considering the appropriate principal submatrix of $A$. This is true also for our main result in Theorem~\ref{thm:main-full}.

\begin{remark} \label{rem:spectral-influence}
    We write the condition on pinnings in Theorem~\ref{thm:main-prev-Bochner} (and in Theorem~\ref{thm:main-full} below) in terms of the spectral gap of the two-site Glauber dynamics, but one can also write it in terms of the associated bipartite random walk operator; see Corollary~\ref{cor:main-hdx-corr}. The equivalent bound on the second eigenvalue in this case becomes an upper bound by $a_{uv}$. This type of bound is typical in the high-dimensional expanders literature, when the goal is to prove local spectral expansion. This is also equivalent to bounding the \textit{spectral influence} between $u$ and $v$ by $a_{uv}$. This quantity was originally defined in \cite{LO25} to obtain trickle-down theorems for finite spins systems. This quantity was also used in \cite{GZ26} in the form of maximal correlations (see Remark~2 of that paper). We discuss the connections between these and other quantities further in Section~\ref{subsec:spectral-gaps}.
\end{remark}

\subsection{Main result: Spectral gaps for block Glauber dynamics}
Our main technical result is then an extension of
Theorem~\ref{thm:main-prev-Bochner} to bounds on spectral gaps of
marginal distributions on disjoint sets of sites. To state this result,
we need some new notation. Let $T_1,\ldots,T_m$ be disjoint sets of
sites and let $T=T_1\sqcup\cdots\sqcup T_m$. For each $i\in[m]$,
write $Y_i=X_{T_i}$ for the tuple of spins on $T_i$, with state space
$\mathcal S_i=\prod_{v\in T_i}S_v$.

Consider the joint law of $(Y_1,\ldots,Y_m)$ on
$\mathcal S_1\times\cdots\times\mathcal S_m$, which is the marginal
of $\mu$ on the sites in $T$. This can be considered as an $m$-site
spin system, where vertex $i$ has spin space $\mathcal S_i$.
We can now consider the associated Glauber dynamics, which chooses
a block uniformly and resamples its configuration conditioned on
the other blocks under this marginal law. As above, we define $P_{\mathrm{Glaub}}^{T_1,\ldots,T_m}$ and $L_{\mathrm{Glaub}}^{T_1,\ldots,T_m}$ for a given tuple of disjoint subsets $T_1,\ldots,T_m$ of $V$.

\begin{theorem} \label{thm:main-full}
    Fix a spin system on $G=(V,E)$ with configurations distributed according to $\mu$, such that $\ker(L_{\mathrm{Glaub}})$ consists only of constant functions (connectedness). Fix also a $V \times V$ symmetric matrix $A$ with non-negative entries and zero diagonal such that $\lambda_{\max}(A) < 1$. Define $M = I-A$, and for any disjoint subsets of sites $T_1,\ldots,T_m \subset V$ with $T = \sqcup_i T_i$ define
    \[
        C = ((M^{-1})_T)^{-1}, \quad D = \operatorname{blockdiag}(C_{T_1}, \ldots, C_{T_m}), \quad H = I - D^{-1/2}CD^{-1/2},
    \]
    where $M_\alpha$ denotes the principal submatrix of $M$ with rows and columns $\alpha$. If for all $u \neq v$ in $V$ and all pinnings $\tau$ of $V \setminus \{u,v\}$ we have
    \[
        \lambda_2(L_{\mathrm{Glaub}}^\tau) \geq \frac{1-a_{uv}}{2},
    \]
    then
    \[
        \lambda_2(L_{\mathrm{Glaub}}^{T_1,\ldots,T_m}) \geq \frac{1-\lambda_{\max}(H)}{m}. % \geq \frac{1-\lambda_{\max}(A)}{m}.
    \]
\end{theorem}

To see how this relates to Theorem~\ref{thm:main-prev-Bochner}, let $T_1,\ldots,T_n$ be singleton sets which completely partition the sites of the spin system, so that $L_{\mathrm{Glaub}}^{T_1,\ldots,T_n} = L_{\mathrm{Glaub}}$, the usual Glauber dynamics on configurations. In this case $C = I-A$, $D = I$, and $H = A$, and therefore Theorem~\ref{thm:main-full} recovers Theorem~\ref{thm:main-prev-Bochner}.

The dependence on the full matrix $A$ allows these bounds to exploit the structure of the conditional interactions when sites are discarded or grouped. We discuss consequences for path complexes, $\lambda$-products, and swap walks in Section~\ref{sec:intro-hdx}.

\begin{remark}
    We note that Theorem~\ref{thm:main-full} is tight for every admissible matrix $A$ and every choice of disjoint blocks. In the notation of Theorem~\ref{thm:main-full}, let $(X_1,\ldots,x_n) = X \sim N(0,M^{-1})$ be a Gaussian random vector, and consider it as a spin system with sites $1,2,\ldots,n$ and spins in $\R$ given by the entries of $X$. Then the hypotheses of Theorem~\ref{thm:main-full} are satisfied with equality, and so are the conclusions for any choice of disjoint $T_1,\ldots,T_m$. We omit the details here.
\end{remark}

\subsection{Application: The $O(N)$ model} \label{sec:main-ON}

We can now give our first application, on a global spectral gap and sampling of the $O(N)$ model.

We first describe the $O(N)$ model. Let $G=(V,E)$ be a finite connected graph with $n=|V|$,
and associate to each vertex $x\in V$ a spin
$\sigma_x\in\mathbb S^{N-1}\subset\mathbb R^N$.
The interactions are specified by a symmetric matrix
$J=(J_{xy})_{x,y\in V}$, where $J_{xx}=0$ and $J_{xy}=0$
unless $x\sim y$. Given external fields $h_x\in\mathbb R^N$,
the associated $O(N)$ Gibbs measure is
\begin{equation}\label{eq:on-model}
\mu_{J,h}(d\sigma)
\propto \exp\left(
\frac12\sum_{x,y\in V}J_{xy}\sigma_x\cdot\sigma_y
+\sum_{x\in V}h_x\cdot\sigma_x
\right)\prod_{x\in V}d\omega(\sigma_x),
\end{equation}
where $\omega$ is the uniform probability measure on
$\mathbb S^{N-1}$. We define
\[
    J_{\mathrm{abs}}=(|J_{xy}|)_{x,y\in V} \quad\text{and}\quad
    R=\max_{x\in V}\sum_{y\ne x}|J_{xy}|.
\]

For this model, some spectral gap bounds are known when the interactions between sites are bounded above in a suitable sense.
The Brascamp-Lieb argument of Dyson, Lieb, and Simon
\cite[Appendix~D]{DLS78} gives a bound on the covariance matrix
of $(\sigma_x)_{x\in V}$ when $\lambda_{\max}(J)-\lambda_{\min}(J)<N$. In particular, this condition holds when $R<N/2$.
Bauerschmidt and Bodineau \cite{BB19} strengthened this to a logarithmic Sobolev inequality for Langevin dynamics
under the same spectral condition. More recently, Anari,
Koehler, and Vuong
\cite[Theorems~5 and~8, Section~1.3]{AKV24} proved a 
logarithmic Sobolev inequality for Langevin dynamics and
$O(n\log n)$ mixing of Glauber dynamics throughout the larger
range $R<(4^{1/3}-1)N\approx0.5874N$.

We now give our first main result, which establishes a spectral
gap of order $1/n$ and thereby $O(n^2)$ mixing time for Glauber dynamics
throughout the range $R<N$. More generally, our result applies whenever
$\norm{J_{\mathrm{abs}}}_{\op}<N$, since $\norm{J_{\mathrm{abs}}}_{\op}\le R$.
We comment that the constant $N$ in this row-sum criterion is sharp for a
uniform spectral gap of order $1/n$, as shown via analysis of phase transitions in the mean-field model \cite{KN16}.

\begin{restatable}{theorem}{maingap}\label{thm:main-gap}
Let $N\ge2$, and let $\mu$ be the $O(N)$ Gibbs measure
\eqref{eq:on-model} on a finite nonempty graph with arbitrary
external fields. Then the spectral gap of Glauber dynamics  is at least
\[
    \frac{1-\frac{\norm{J_{\mathrm{abs}}}_{\op}}{N}}{n}.
\]
Consequently, for every $\varepsilon_0\in(0,1)$, if
$\norm{J_{\mathrm{abs}}}_{\op}\le(1-\varepsilon_0)N$, then Glauber dynamics mixes
in $O_{N,\varepsilon_0}(n^2)$, uniformly
in the external fields.
\end{restatable}

Using established tools, we can improve this mixing-time bound to the
optimal $O(n\log n)$; see Remark~\ref{rem:nlogn}. The local 2-link bound used to prove Theorem~\ref{thm:main-gap}
also yields bounds on marginal correlations and block dynamics,
which we discuss below (see Remark~\ref{rem:marginal-corr}).

\subsection{Applications in high-dimensional expanders} \label{sec:intro-hdx}

Theorem~\ref{thm:main-full} has a number of connections and applications to spectral bounds studied in the field of high-dimensional expanders (HDX). In the context of HDX, spin systems are referred to as partite simplicial complexes: the sites specify vertices of different types, configurations are facets which contain one vertex of each type, and pinnings correspond to passing to links. Bounding the spectral gap of the Glauber dynamics on various choices of block of sites (as allowed by Theorem~\ref{thm:main-full}) is then equivalent to different notions of expansion of the given simplicial complex.

\paragraph{$\lambda$-products and swap walks.}
One such notion is that of a $\lambda$-product, introduced by \cite{GLL22} and studied later in \cite{DH24,Hop25}. A distribution $\mu$ is a \textbf{$\lambda$-product} if, after any pinning, the maximal correlation $\rho_{\max}$ between any two remaining coordinates is at most $\lambda$. (This is equivalent to bounds on the spectral gaps of the associated pinned Glauber dynamics, see the definition of $\rho_{\max} = \rho_\mu$ and related results in Section~\ref{subsec:spectral-gaps}.) This condition then controls correlations (and hence also the Glauber spectral gap) between groups of coordinates: for disjoint sets $T_1,T_2$, Lemma~3.3 of \cite{GLL22} gives
\[
    \rho_{\max}(X_{T_1}, X_{T_2}) \leq \lambda \sqrt{|T_1| \, |T_2|},
\]
where $X \sim \mu$ and $\rho_{\max}$ denotes the maximal correlation. This quantity also bounds the spectral gap of the associated \textbf{swap walk}, which is the bipartite random walk on the spins of the groupings $T_1$ and $T_2$. These walks were originally studied in \cite{DD19,AJT19} and later in \cite{DD24,AP26}.

By the two-site equivalence in Lemma~\ref{lem:equiv-forms}, the two-block case of Theorem~\ref{thm:main-full} gives
\[
    \rho_{\max}(X_{T_1},X_{T_2}) \leq \lambda_{\max}(H) = \left\|C_{T_1}^{-1/2}C_{T_1,T_2}C_{T_2}^{-1/2}\right\|_{\mathrm{op}}.
\]
This estimate retains the locations and strengths of the conditional interactions recorded in $A$, allowing it to exploit sparse interaction patterns. Moreover, $\lambda_{\max}(H) \leq \lambda_{\max}(A)$, giving a bound independent of the sizes of $T_1$ and $T_2$.

\paragraph{Path complexes and flags of subspaces.}

For the path complexes studied in \cite{LLO25}, we can take $A=\frac12\operatorname{Adj}(P_n)$, where $P_n$ is the path on $n$ vertices. Computing $(I-A)^{-1}$ and applying the two-singleton case of Theorem~\ref{thm:main-full} gives
\[
    \rho_{\max}(X_i,X_j) \leq \lambda_{\max}(H) = \sqrt{\frac{i(n+1-j)}{j(n+1-i)}}, \quad i<j.
\]
This recovers Theorem~7.2 of \cite{LLO25}, while Theorem~\ref{thm:main-full} also applies to arbitrary disjoint groups of sites.

A related example regards chains/flags of subspaces in a finite vector space. Consider a uniformly random complete flag $0 \subset V_1 \subset \cdots \subset V_{d-1} \subset \mathbb{F}_q^d$. Conditioned on the other ranks, nonadjacent coordinates are independent. And after pinning everything else, the distribution on two adjacent coordinates reduces to the lines and planes distribution of $\mathbb{F}_q^3$. A standard computation implies that we can use
\[
    A = \frac{\sqrt{q}}{q+1} \operatorname{Adj}(P_{d-1}), \quad \text{for which} \quad \lambda_{\max}(A) < \frac{2\sqrt{q}}{q+1}.
\]
For singleton sites/rank sets, applying Theorem~\ref{thm:main-full} implies
\[
    \rho_{\max}(V_i,V_j) \leq q^{-|i-j|/2},
\]
and the same holds for all pinnings. This recovers the $q^{-1/2}$-product property studied in \cite{HR26} for flags of subspaces.

Theorem~\ref{thm:main-full} also allows for larger groups of ranks/sites, and the upper bound described above implies
\[
    \rho_{\max}\big((V_i)_{i \in T_1}, (V_j)_{j \in T_2}\big) \leq \lambda_{\max}(H) < \frac{2\sqrt{q}}{q+1},
\]
a bound that also holds for the corresponding swap walk. We note that the general $\lambda$-product bound in this case gives an upper bound of $\frac{r+1}{\sqrt{q}}$ for groups of size $r+1$. Our bound above, on the other hand, is independent of $r$.

\paragraph{Local spectral expansion.}
As a further application within HDX, we state a corollary of Theorem~\ref{thm:main-full} for local spectral expansion on simplicial complexes. The proof is essentially immediate once the various equivalent spectral gap notions are unwound, see Section~\ref{subsec:spectral-gaps}.

\begin{corollary} \label{cor:main-hdx-corr}
    Assume the notation and assumptions of Theorem~\ref{thm:main-full}. Assume additionally that Glauber dynamics is connected after every admissible pinning. Let $X$ be a finite weighted $m$-partite complex whose vertices in part $a$ are spin assignments to $T_a$, and whose facet distribution is the joint law of $Y=(X_{T_1},\ldots,X_{T_m})$. Then $X$ is a one-sided $\lambda_{\max}(H)$-local spectral expander.
    
    More precisely, let $\tau$ be a face leaving the parts indexed by $I \subseteq [m]$ unpinned with $|I| \geq 2$, and put $U=\bigcup_{a\in I}T_a$. If $W_\tau$ is the normalized adjacency operator of the one-skeleton of $\mathcal X_\tau$, then
    \[
        \lambda_2(W_\tau) \leq \lambda_{\max}(H_U) \leq \lambda_{\max}(H)<1
    \]
    where $H_U$ is the principal submatrix indexed by $U$.
\end{corollary}

Applying Corollary~\ref{cor:main-hdx-corr} to flags of subspaces gives a consequence for the colored faces complexes studied by Hopkins and Ray \cite{HR26}. Fix a set of ranks/sites $S\subseteq[d-1]$ and pairwise disjoint subsets $T_1,\ldots,T_m\subseteq S$, each of size $r+1$. The associated colored faces complex has partial flags on $T_a$ as its vertices in part $a$, with facet distribution induced by a uniformly random complete flag. For these complexes, Lemma~49 of \cite{HR26} gives the local spectral expansion bound
\[
\frac{2(r+1)}{\sqrt q-1-2(|S|-r)}
\]
in the parameter range where this bound is nontrivial. Corollary~\ref{cor:main-hdx-corr} applies to the same complexes and gives
\[
\lambda_2(W_\tau)
\le \lambda_{\max}(H)
<\frac{2\sqrt q}{q+1}
\]
for every link with at least two remaining parts. Thus, we obtain a one-sided local spectral expansion bound that
depends only on $q$ and is nontrivial for every prime power $q$.

\subsection{Application: Negative correlations} \label{sec:main-neg-corr}

The Bochner framework also yields covariance identities, which we use to study pairwise inclusion probabilities for random forests and matroid independent sets. Concretely let $G = (V,E)$ be a graph, and let $\mu$ be the uniform distribution on subsets of $E$ which are subforests of $G$. A famous conjecture then states a strong negative correlation property of marginal edge probabilities of $\mu$; it claims that for all $e \neq f \in E$ that
\begin{equation} \label{eq:neg-corr}
    c \cdot \Pr_{F \sim \mu}[e \in F] \Pr_{F \sim \mu}[f \in F] \geq \Pr_{F \sim \mu}[e,f \in F] \quad \iff \quad c \cdot \Pr_{F \sim \mu}[f \in F] \geq \Pr_{F \sim \mu}[f \in F \mid e \in F]
\end{equation}
for $c=1$.

Another very related question asks something similar for independent sets and elements of a matroid. A \textbf{matroid} $M$ on ground set $E$ can be defined as a downward-closed collection $\mathcal{I}$ of subsets of $E$ such that for all $A,B \in \mathcal{I}$ such that $|A| < |B|$, there exists $e \in B \setminus A$ that $A \cup \{e\} \in \mathcal{I}$. The distribution $\mu$ we consider is then uniform on $\mathcal{I}$, and it is unknown what the optimal value of $c$ is in this case.
The general log-concavity results of \cite{BH20,ALOV24} imply the bound \(c=2\) for both matroid independent sets and forests.

For matroids, prior work shows that $c \geq \frac{4}{3}$ (see the Divoux-Wang construction in \cite{Eur26MatroidCorrelations} or \cite{HSW22} for a bound of $\frac{8}{7}$). For forests, there are several partial results toward the conjectured
bound $c=1$, including for small graphs
\cite{GW04}, series-parallel graphs \cite{Eri12},
and for weighted forests in certain large-fugacity
regimes \cite{CR26}.

Matroids capture a notion of independence which generalizes linear independence ($E$ is a set of vectors and $\mathcal{I}$ is the set of linearly independent subsets of $E$) and forests of a graph ($E$ is the set of edges of a graph, and $\mathcal{I}$ is the set of acyclic subsets of $E$, as in the forest example above). We note that these examples can also be seen as spin systems: $E$ is the set of sites, and possible spins are either $1$ or $0$ depending on if a given $e \in E$ is included in the random set/forest or not.

We now show that \(c=\frac{4}{3}\) holds for uniform random forests, improving the factor \(2\) bound supplied by the general results above.

\begin{theorem} \label{thm:main-neg-corr-forests}
    If $\mu$ is the uniform distribution on subforests of a graph, then $c = \frac{4}{3}$ holds in \eqref{eq:neg-corr}.
\end{theorem}

The covariance identity underlying our argument also gives a proof of the factor $2$ bound for matroid independent sets, including under arbitrary external fields. For forests, an auxiliary real-stable polynomial gives finer control of the same covariance entries and yields the improvement to $\frac{4}{3}$; see Section~\ref{subsec:forests}.

\subsection{Technical Overview} \label{sec:tech-overview}

We now give a brief overview of the structure and proofs of our results. We first describe the discrete spins case, which is easier to think about. Later we will discuss how things change to generalize to continuous spins, which for example is needed to handle the $O(N)$ model. In this case, $P_{\mathrm{Glaub}}$ is a $(\# \text{ of configurations})$-size square matrix with entries computing the random walk transitions from one configuration to another. This matrix can thus be viewed as a linear operator on functions $f: \{\text{configurations}\} \to \R$, and we will adopt this mentality throughout. Recall that our overarching goal is to bound the spectral gap of $P_{\mathrm{Glaub}}$.

\paragraph{Bochner and Poincar\'e inequalities.}

The first basic observation is that, for a given site $v \in V$, the probability transition matrix restricted to that site (i.e., once $v$ has been selected as the site to update) is a conditional expectation operator. Concretely, we take expectation conditioned on fixing all spins except for the spin at site $v$. Letting $X = (X_v)_{v \in V}$ be a random variable associated to $\mu$, this operator is given by $E_v f = \E_\mu[f \mid X_{-v}]$ where $X_{-v}$ denotes the values of $X_u$ in $f$ for all $u \neq v$. This also gives us a concrete description of the transition matrix for the Glauber dynamics: simply average over the sites to obtain $P_{\mathrm{Glaub}} = \frac{1}{|V|} \sum_{v \in V} E_v$.

Now let us look at the local Laplacians, given by $L_v = I - E_v$. By the formula above, we obtain the expression $L_{\mathrm{Glaub}} = I - P_{\mathrm{Glaub}} = \frac{1}{|V|} \sum_{v \in V} L_v$ for the global Laplacian. Since $P_{\mathrm{Glaub}}$ is a probability transition matrix and thus has maximum eigenvalue $1$, the desired spectral gap becomes a bound of the form $L_{\mathrm{Glaub}}^2 \succeq \frac{\epsilon}{|V|} L_{\mathrm{Glaub}}$ for a spectral gap of $\epsilon/|V| > 0$. (Note that this assumes the graph of $P_{\mathrm{Glaub}}$ is connected, or else $L_{\mathrm{Glaub}}$ may have multiple $0$ eigenvalues.) The Bochner method then relates $L_{\mathrm{Glaub}}$ and $L_{\mathrm{Glaub}}^2$ via the local Laplacians using the decomposition $L_{\mathrm{Glaub}} = \frac{1}{|V|} \sum_{v \in V} L_v$, implying a spectral gap. Thus we want a bound of the form
\begin{equation} \label{eq:Bochner-tech-overview}
    \big\|L_{\mathrm{Glaub}} f\big\|^2 = \left\|\frac{1}{|V|} \sum_{v \in V} L_v f\right\|^2 \geq \frac{\epsilon}{|V|^2} \sum_{v \in V} \|L_v f\|^2 = \frac{\epsilon}{|V|} \langle f, L_{\mathrm{Glaub}} f \rangle
\end{equation}
for all functions $f$.

A further equivalent formulation is via a Poincar\'e inequality, which compares the total variance of a given function $f$ to the sum of its local conditional variances at each site. The local conditional variances are captured by the local Laplacians, since $\|L_v f\|^2_2 = \E_\mu[\Var(f \mid X_{-v})]$ for all $v \in V$, and the total variance is given by $\Var_\mu(f) = \|(I - \E_\mu)f\|^2$, where $\E_\mu$ is the global expectation operator. The Poincar\'e inequality equivalent to the above Bochner inequality \eqref{eq:Bochner-tech-overview} is
\begin{equation} \label{eq:Poincare-tech-overview}
    \Var_\mu(f) = \|(I - \E_\mu)f\|^2 \leq \epsilon^{-1} \sum_{v \in V} \|L_v f\|^2_2 = \epsilon^{-1} \sum_{v \in V} \E_\mu[\Var(f \mid X_{-v})]
\end{equation}
for all functions $f$.

\paragraph{The inequalities via linear algebra.}

The inequalities \eqref{eq:Bochner-tech-overview} and
\eqref{eq:Poincare-tech-overview} can be generalized to purely linear
algebraic statements about orthogonal projections.
We let $\mathcal H=L^2(\mu)$, with inner product
$\langle f,g\rangle=\sum_\sigma\mu(\sigma)f(\sigma)g(\sigma)$.
Then $L_v$ is an orthogonal projection for all $v\in V$, and
$I-\E_\mu$ is also an orthogonal projection onto the space of
mean-zero functions. Equivalently, $I-\E_\mu$ is the orthogonal projection onto
$\overline{\sum_{v\in V}\operatorname{ran}(L_v)}$.
Indeed, connectedness gives
\[
    \left(\sum_{v\in V}\operatorname{ran}(L_v)\right)^\perp
    =\bigcap_{v\in V}\ker L_v
    =\operatorname{span}\{1\}.
\]

We first state the abstract inequalities that give the global bound.
For spin systems, this corresponds to Theorem~\ref{thm:main-prev-Bochner}, or the case of
Theorem~\ref{thm:main-full} where the blocks are singleton sets
covering all sites.

\begin{proposition}[Bochner and Poincar\'e inequalities]
    \label{prop:1-2-tech-overview}
    Let $P_i$ be orthogonal projections on a Hilbert space
    $\mathcal H$ with images given by $W_i$ for all $i\in[n]$,
    and let $S$ be a symmetric $n\times n$ matrix with non-negative
    entries, zero diagonal and maximum eigenvalue
    bounded by $1-\epsilon$ where $\epsilon>0$.
    Define $P_{ij}$ and $P$ to be the orthogonal projections onto
    $W_i\cap W_j$ and
    $\overline{\sum_{i=1}^nW_i}$, respectively.
    If $\|P_iP_j-P_{ij}\|\leq s_{ij}$ for all $i\ne j$, then
    \[\left\|\sum_{i=1}^nP_if\right\|^2 \geq\epsilon\sum_{i=1}^n\|P_if\|^2
        \quad\text{and}\quad
        \|Pf\|^2\leq\epsilon^{-1}\sum_{i=1}^n\|P_if\|^2
    \]
    for all $f\in\mathcal H$.
\end{proposition}

One interpretation of the condition
$\|P_iP_j-P_{ij}\|\leq s_{ij}$ is that the angle between vectors in
$W_i$ and $W_j$, orthogonal to their intersection, cannot be too
small; see Lemma~\ref{lem:angle-proj-norm}. More importantly for spin systems, this condition follows from a
two-site Glauber spectral-gap bound of $(1-s_{ij})/2$, uniformly
over pinnings of the other sites, which is how the local assumptions of
Theorem~\ref{thm:main-prev-Bochner} and Theorem~\ref{thm:main-full} enter the argument.

We now give the main ideas of the proof. We consider the following sequence of maps between
Hilbert spaces:
\[
    \mathcal H_0\xrightarrow{d_0}\mathcal H_1
    \xrightarrow{d_1}\mathcal H_2,
\]
where
\[
    \mathcal H_0=\mathcal H,\qquad
    \mathcal H_1=\bigoplus_iW_i,\qquad
    \mathcal H_2=\bigoplus_{i<j}W_{ij},
\]
with the direct-sum inner products inherited from
$\mathcal H$, and
\[
    d_0:f\mapsto(P_if)_i
    \quad\text{and}\quad
    d_1:(f_i)_i\mapsto
        \big(P_{ij}(f_i-f_j)\big)_{i<j}.
\]
The proof of Proposition~\ref{prop:1-2-tech-overview} then follows
from analyzing a single operator $B:\mathcal H_1\to\mathcal H_1$,
defined by $B=d_0d_0^*+d_1^*d_1$.
We refer to this as the \textbf{Bochner operator}.
One of the key facts about this operator is that its off-diagonal
$W_j \to W_i$ blocks are precisely
$P_iP_j-P_{ij}$, allowing us to use the norm bound assumption.

To see how the inequalities follow, we first observe that
\[
    d_0^*:(f_i)_i\mapsto\sum_i f_i
    \quad\implies\quad
    d_0^*d_0=\sum_iP_i.
\]
Our choice of $d_1$ gives $d_1d_0=0$. Using this, we can relate
$B$ and $B^{-1}$, when the inverse exists, to $\sum_iP_i$ and $P$:
\[
    \left(\sum_iP_i\right)^2=d_0^*Bd_0
    \quad\text{and}\quad
    P=d_0^*B^{-1}d_0.
\]

These are the identities underlying Lemma~\ref{lem:Bochner-lite}.
We are therefore left with proving $B\succeq\epsilon I$.
The diagonal blocks satisfy $B_{ii}\succeq I$, while the
off-diagonal blocks satisfy $\|B_{ij}\|\leq s_{ij}$.
Together with $\lambda_{\max}(S)\leq1-\epsilon$, these give
$B\succeq\epsilon I$, and hence
$B^{-1}\preceq\epsilon^{-1}I$.
The identities above now imply
\[
    \left(\sum_iP_i\right)^2
    \succeq\epsilon\sum_iP_i
    \quad\text{and}\quad
    P\preceq\epsilon^{-1}\sum_iP_i.
\]
Taking quadratic forms and using the projection identities $P_i^2=P_i$ and $P^2=P$ gives
the desired inequalities. For a more detailed proof, see
Section~\ref{sec:proof-abstract-proj}.

\paragraph{Finishing the proof of Theorem~\ref{thm:main-full}.}

The above argument handles Theorem~\ref{thm:main-prev-Bochner}, but how does it extend to capture all of Theorem~\ref{thm:main-full}?

We give a brief overview here, leaving the full proof to Section~\ref{sec:proof-abstract-proj}.

Recall for the general result Theorem~\ref{thm:main-full}, we want to analyze a collection of disjoint subsets of sites $T_1,\ldots,T_m$. The next step at this point towards finishing the proof is to somehow restrict $T = T_1 \sqcup \cdots \sqcup T_m$ to continue the analysis, where each $T_a$ groups some disjoint set of the indices $i \in [n]$. In the notation here, we thus decompose $\mathcal{H}_1 = \mathcal{H}_1^T \oplus \mathcal{H}_1^R$ in the natural way:
\[
    \mathcal{H}_1^T = \bigoplus_{i \in T} W_i.
\]
For a function $f$ of the retained spins indexed by $T$, the components of $d_0 f$ outside $T$ vanish. Since $P = d_0^* B^{-1} d_0$, the quantity $\|Pf\|^2$ is thus expressed via the compression $(B^{-1})_{TT}$, so the reduced operator to study is
\[
    \tilde{B} = \bigl((B^{-1})_{TT}\bigr)^{-1}.
\]
This is the Schur complement of $B$, just as $C$ is the corresponding Schur complement of $M$. The benefit of Schur complement over something simpler like projection is that it retains the effect of the discarded sites on interactions among the sites that remain.

Once we take Schur complement, we now have this ``reduced'' Bochner operator $\tilde{B}: \mathcal{H}^T_1 \to \mathcal{H}^T_1$, the same coercivity $\epsilon I \preceq \tilde{B}$ follows from a standard lemma (Lemma~\ref{lem:compression-schur}). To complete the block-wise analysis on $T_1,\ldots,T_m$ though, we need a more refined coercivity bound. Concretely, we prove
\[
    \tilde{B} \succeq \alpha \tilde{D} \quad \text{where} \quad \tilde{D} = \operatorname{blockdiag}(\tilde{B}_{T_1T_1}, \ldots, \tilde{B}_{T_mT_m}),
\]
and where $\alpha = \lambda_{\min}(D^{-1/2}CD^{-1/2})$ is the numerator of the desired bound in Theorem~\ref{thm:main-full} (with $D = \operatorname{blockdiag}(C_{T_1T_1},\ldots,C_{T_mT_m})$).

We then bring everything back to a statement about orthogonal projections. We consider projections $\tilde{Q}_a$ in $\mathcal{H}$ which combine both the Schur complementing to $T$ and also the grouping of the indices $i \in T_a$ for each $a \in [m]$. Similarly we consider $\tilde{P}$, which is the analogous version of the projection $P$ onto $W$. With further analysis, using the coercivity proven above, we then obtain the bound
\[
    \sum_{a=1}^m \|\tilde{Q}_a f\|^2 \geq \alpha \|\tilde{P} f\|^2.
\]
This is the correct analogue of Proposition~\ref{prop:1-2-tech-overview} in this more general context which leads to the conclusion of Theorem~\ref{thm:main-full}. See Section~\ref{sec:proof-of-main-full} for the final details which make the connection back to the Glauber dynamics.

We finally comment that the full proofs of all these results require working on a possibly infinite dimensional Hilbert space in order to be able to handle infinite spin spaces for spin systems. Everything we discussed above immediately generalizes to this case, with the right (but standard) care taken.

The remainder of this overview is spent discussing our main applications.

\paragraph{The $O(N)$ model.}
As we have already observed, the Bochner identity reduces the spectral-gap bound to controlling the
cross terms $\langle L_xf,L_yf\rangle$. Each of these resulting two-links has a
conditional distribution of the form $\nu(du,dv)\propto
e^{a\cdot u+b\cdot v+\kappa u\cdot v}\,d\omega(u)\,d\omega(v)$. By Theorem~\ref{thm:main-full}, it suffices to show that, in every
codimension-$2$ link, the maximal correlation between $U$ and $V$
is at most $|\kappa|/N$, uniformly in $a$ and $b$. Using the random walk interpretation on the link (see Section~\ref{sec:random-walk}), this can be proved by showing that for every Lipschitz $\varphi$ and $(T \varphi)(u)=\mathbb E[\varphi (V)\mid U=u] $, we have $|T\varphi (u)-T \varphi(u')|
\leq \frac{|\kappa|}{N}\,\ell\|u-u'\|$ (see the proof of Lemma~\ref{lem:two-spin} in Section~\ref{subsec:proof-local}). 

Since changing $u$ changes only the external field seen
by $V$, this is equivalent to understanding how $\varphi(V)$ changes under varying external fields. To that end, we observe that \[\frac{d}{ds}\E_{\nu_{h+sc}}\varphi(U)
=\Cov_{\nu_{h+sc}}(\varphi(U),c\cdot U)\] and so we are reduced to upper-bounding this covariance quantity by $\norm c\Lip(\varphi)/N$.

\emph{The case of $h=0$.} Since
$\mathbb E_{\nu_0} U=0$, we write using integration by parts
\begin{align*}
|\operatorname{Cov}_{\nu_0}(\varphi(U),c\cdot U)|
&=\Big|\mathbb E_\omega[\varphi(U)c\cdot U]\Big| =\frac1N
  \Big|\mathbb E_{Z\sim\mathrm{Unif}(\mathbb B^N)}
  [c\cdot\nabla F(Z)]\Big| \leq\frac{\|c\|}{N}\operatorname{Lip}(\varphi).
\end{align*}
Here $c\in\mathbb R^N$ is arbitrary, and $F$ is a Lipschitz
extension of $\varphi$ with the same Lipschitz constant, as provided
by Proposition~\ref{prop:whitney-extension}.

\emph{General $h$.} By rotational invariance we may assume that $h = te$, where $e$ is any coordinate vector. To generalize the $h=0$ computation above  we seek a vector field $V_c$ with
\[
\nabla\cdot V_c=0 \quad\text{and} \quad V_c(u)\cdot u=w(u)\,c\cdot(u-m_t e) \quad\text{where } w(z)=\frac{e^{te\cdot z}}{\mathbb E_\omega[e^{te\cdot U}]}.
\]
Integrating by parts (or precisely, the divergence theorem) would give the following analogous expression to the above
\[
\operatorname{Cov}_{\nu_h}(\varphi(U),c\cdot U)
=\frac1N\,\mathbb E_{Z \sim \mathrm{Unif}(\mathbb B^N)}[\nabla F(Z)\cdot V_c(Z)].
\]

To that end, we split $c$ into radial and tangential directions relative to $e$. For the tangential direction $c=v\perp e$, the tilt has no gradient in the direction of integration and so this case is essentially the same as $h=0$ and we can take $V_v = wv$. The radial direction $c=e$ requires more work: we need a field $V$
with zero divergence and boundary flux $w(u)(e\cdot u-m_t)$,
whereas $\nabla\cdot(we)=tw$. We instead solve the boundary-value problem
\[
\Delta A=t^2A \quad\text{in }\mathbb B^N,
\qquad
A=1 \quad\text{on }\mathbb S^{N-1}.
\]
The idea is that $w$ satisfies the same equation, $\Delta w=t^2w$. For $A$ satisfying the above, we may then define $V_e=\frac{A\nabla w-w\nabla A}{t}$ which then satisfies the correct boundary conditions on $\mathbb{S}^{N-1}$ and is divergence-free since $\nabla\cdot V_e=\frac{A\Delta w-w\Delta A}{t}=0$.

\paragraph{Negative correlations in matroids.}
We next consider the problem of showing negative correlations for random
independent sets from a matroid. Here, it is useful to think about what
$L_iL_j-L_{ij}$ measures. 
$L_iL_j -L_{ij}$ vanishes when the two coordinates are conditionally
independent, so we can think of it as recording their local interaction. It is then
perhaps not too surprising that $B^{-1}$ recovers correlations; see
Lemma~\ref{lem:Bochner-lite} and its proof for the precise formulation.

Having this, to recover the bound $p_{ij} \leq 2 p_ip_j$ where $p_i$ is the probability that $i$ is in the random independent set and $p_{ij}$ that both $i,j$ are in the random independent set, we study $B^{-1}$.

For this application, we use the weighted inner products introduced in Section~\ref{subsec:neg-corr-notation}. The Bochner operator still has the form $B=d_0d_0^*+d_1^*d_1$, with adjoints taken with respect to these inner products. The main step is to prove $B\succeq I$, and hence $B^{-1}\preceq I$.

Assuming this, we use Lemma~\ref{lem:Bochner-lite} which shows that
$\Sigma\preceq\operatorname{diag}(p)$ for the covariance matrix $\Sigma$ and we write $\Delta=\operatorname{diag}(p)-\Sigma\succeq0$.
Since $\Delta_{ii}=p_i^2$, this immediately implies the desired conclusion: for $i\neq j$, its $2\times2$ principal minors give
\begin{equation}\label{eq:negative-matroid}
    p_{ij}
    =p_ip_j-\Delta_{ij}
    \le p_ip_j+\sqrt{\Delta_{ii}\Delta_{jj}}
    =2p_ip_j.
\end{equation}

It remains to justify why $B\succeq I$. This is what we do in
Proposition~\ref{prop:matroid-bochner-inverse}. To understand the off-diagonal entries, we condition on all coordinates except $i,j$. If
$S\cup\{i,j\}$ is independent, where $S$ is the remaining set, then
the conditional law is a product  measure and the off-diagonal interaction
vanishes. The only interesting case is when we can add either element
separately but cannot add both. For matroids, this means exactly that
$i,j$ belong to the same parallel class of $M/S$. The key identity,
\eqref{eq:matroid-main}, shows that within each class these
cross terms combine with the corresponding diagonal terms
into a nonnegative square. Summing these up then gives the desired $B\succeq I$.

\smallskip
\emph{Graphic matroids.}
We next consider forests, where we can improve the factor $2$ to $4/3$
in \eqref{eq:negative-matroid}. To that end, we need a more careful
analysis that uses the graphic matroid structure. As a starting point, we
recall that for spanning trees, one way to prove negative correlation
is from the Rayleigh inequalities after showing real stability of the spanning-tree polynomial. We can imitate such an approach in the case of forests as well.

To obtain a
related polynomial construction for forests, we add an auxiliary vertex and connect
it to each vertex $v$ by an edge of weight $x_v$, giving a new graph $G'$.
If we delete the auxiliary vertex from a spanning tree of $G'$, then
we obtain a forest with one chosen root in each component. We also label the original edges with the variables $z_e$ and assign edge $e$ the corresponding
weight $\lambda_e(1+z_e)$. We will effectively work with the spanning-tree polynomial of this weighted graph.

The idea is that differentiating in $z_e$
selects forests containing $e$, while differentiating in two variables
selects forests containing both edges. That is, these variables let us
recover the inclusion probabilities we want to compare. Roughly speaking, this is the motivation behind the 
auxiliary polynomial $\Phi$ (see \eqref{eq:component-matrix} and the proof of Proposition~\ref{prop:component-stability}). 

The main fact we use, proved in
Proposition~\ref{prop:component-stability}, is that $\Phi$ is real stable.
The Rayleigh inequality applied to $\Phi$ then gives
$3p_{ij}\le (2p_i)(2p_j)$, which rearranges to the desired bound.
Through the lens of the covariance identity (see \eqref{eq:matroid-inverse-covariance}), we can view this as a more precise
estimate on the pairings involving $B^{-1}$ that represent pairwise
correlations and the real stability of $\Phi$ encapsulates the additional
graphic structure that enables the improved bound.

\subsection{Acknowledgments}

We would like to thank Shayan Oveis Gharan for important discussions and ideas in the first stages of this paper, when the application to the $O(N)$ model was being developed, and also for the idea to try to connect our results to negative correlations for forests. We would also like to thank Yotam Dikstein and Max Hopkins for conversations which led to the connections to swap walks and the other high-dimensional expansion results. These discussions were also important for the final version of Theorem~\ref{thm:main-full}. We would also like to thank Mathews Boban, Kuikui Liu, Xiaoyu Chen, and Lap Chi Lau for stimulating conversations around the topics of this paper.

AL is supported by the Computer Science \& Engineering Educators' Endowed Fellowship and the Jonathan Hsieh and Kathleen Ting Endowed Fund. JL acknowledges the support of the Natural Sciences and Engineering Research Council of Canada (NSERC), [funding reference number RGPIN-2023-03726]. Cette recherche a \'et\'e partiellement financ\'ee par le Conseil de recherches en sciences naturelles et en g\'enie du Canada (CRSNG), [num\'ero de r\'ef\'erence RGPIN-2023-03726].

The development of this framework and its applications involved substantive assistance from OpenAI’s ChatGPT. This included assistance with the Hilbert-space formulation and the proof of the general block theorem, the real-stability approach to the forest correlation bound, and the radial construction used in the covariance inequality for tilted spherical measures. We describe these contributions and our use of the tool in more detail in the AI methodology statement at the end of the paper.

\section{Preliminaries}\label{sec:preliminaries}

We collect here some standard facts that will be used later.

\subsection{Linear algebra and Hilbert spaces}\label{subsec:linear-algebra}

Throughout, we let $\mathcal H$ denote a real Hilbert space, and we write
$\|A\|_{\mathrm{op}}$ for the operator norm of a bounded operator $A$.
For bounded
self-adjoint operators, we write $A\preceq B$ if
$\langle f,Af\rangle\le\langle f,Bf\rangle$ for all $f\in\mathcal H$.
If $U\subseteq\mathcal H$ is closed, then let $P_U$ denote the orthogonal projection onto $U$. We also define the \emph{compression} of $A$ to $U$ as $A_{UU}=P_UA|_U$. 
Note that when the subspace is a span of standard basis vectors, then the compression simply becomes the corresponding principal submatrix of $A$.

We collect the following standard facts about compressions and Schur
complements of uniformly positive bounded operators; see
\cite{AT75} and
\cite[Section~4]{CMS01}.

\begin{lemma}[Compression and Schur complements]
\label{lem:compression-schur}
Let $\mathcal H=U\oplus V$ be an orthogonal decomposition into closed
subspaces, and let $A$ be bounded and self-adjoint with
$0<aI\preceq A\preceq bI$. Write
\[
    A=\begin{pmatrix}
        A_{UU}&A_{UV}\\ A_{VU}&A_{VV}
      \end{pmatrix}.
\]
Then the following hold.
\begin{enumerate}
    \item The compressions $A_{UU}$ and $(A^{-1})_{UU}$ are boundedly
    invertible, with
    \[
        aI_U\preceq A_{UU}\preceq bI_U,\qquad
        b^{-1}I_U\preceq(A^{-1})_{UU}\preceq a^{-1}I_U.
    \]
    \item The operator $K_U=((A^{-1})_{UU})^{-1}$ satisfies
    \[
        K_U=A_{UU}-A_{UV}A_{VV}^{-1}A_{VU},
        \qquad
        aI_U\preceq K_U\preceq A_{UU}\preceq bI_U.
    \]
    If $V=\{0\}$, the subtracted term is omitted.
    \item For every $x\in U$,
    \[
        \min_{\substack{z\in\mathcal H\\P_Uz=x}}
             \langle z,A^{-1}z\rangle
        =\langle x,A_{UU}^{-1}x\rangle.
    \]
    In particular,
    $\langle z,A^{-1}z\rangle
      \ge\langle P_Uz,A_{UU}^{-1}P_Uz\rangle$ for every $z\in\mathcal H$.
\end{enumerate}
Moreover, compression preserves operator order, while inversion reverses
operator order for bounded, uniformly positive self-adjoint operators.
\end{lemma}

We next record a standard fact about nonnegative self-adjoint operators that allows us to deduce spectral-gap bounds from the Laplacian bounds in the Bochner method.

\begin{fact}\label{fact:spectral-gap-equivalences}
Let $L$ be a bounded, self-adjoint, nonnegative operator on $\mathcal H$,
and let $\Pi$ be the orthogonal projection onto $(\ker L)^\perp$.
For $\gamma>0$, the following are equivalent:
\[
\begin{gathered}
    \sigma(L)\subseteq\{0\}\cup[\gamma,\infty),\\[2pt]
    \gamma\|\Pi f\|^2\le\langle f,Lf\rangle,
    \qquad
    \gamma\langle f,Lf\rangle\le\|Lf\|^2,
\end{gathered}
\]
where both inequalities hold for every $f\in\mathcal H$.
\end{fact}
\begin{proof}
For $g=L^{1/2}f$, we have
$\langle g,Lg\rangle=\|Lf\|^2$ and
$\|g\|^2=\langle f,Lf\rangle$.
Thus the third condition is equivalent to
$\gamma\|g\|^2\le\langle g,Lg\rangle$ on
$\operatorname{ran}L^{1/2}$.
Since
$\overline{\operatorname{ran}L^{1/2}}=(\ker L)^\perp$,
boundedness of $L$ extends this inequality by continuity to
$(\ker L)^\perp$.
This is the second condition, since
$\langle f,Lf\rangle=\langle\Pi f,L\Pi f\rangle$.
The first and second conditions are equivalent by the spectral theorem.
\end{proof}
\subsubsection{Subspace angles}\label{subsec:angles}

Given two non-zero vectors in a Hilbert space $\mathcal{H}$, there is a natural definition of the angle between then coming from the inner product. Specifically, we can define the cosine of the angle between $f,g \in \mathcal{H}$ via
\[
    \cos(f,g) := \frac{\langle f, g \rangle}{\|f\| \, \|g\|}.
\]
this comes from the usual dot product identity. This can then be generalized to (closed) subspaces in a natural way:
\begin{equation} \label{eq:cos-definition}
    \cos(U,W) = \sup_{\substack{u \in U, w \in W \\ u,w \perp U \cap W}} \frac{\langle u,w \rangle}{\|u\| \, \|w\|},
\end{equation}
where the sup is over non-zero vectors $u,w$. Note here that we ignore the intersection of $U$ and $W$, because the above ratio can be trivially made to be $1$ there (e.g. take $u=w$ in $U \cap W$), which doesn't give any information. This quantity $\cos(U,W)$ is called the cosine of the \emph{Friedrichs angle}~\cite{D95}.

We now state a standard result about how this angle can be interpreted as the norm of a natural operator associated to $U$ and $W$.

\begin{lemma}[Angle-norm equivalence]\label{lem:angle-proj-norm}
Let $P_i,P_j$, and $P_{ij}$ be the orthogonal projections onto closed
subspaces $W_i,W_j$, and $W_i\cap W_j$, respectively. Then
\[
    \cos(W_i,W_j)=\|P_iP_j-P_{ij}\|_{\mathrm{op}}.
\]
\end{lemma}
\begin{proof}
    Note first that
    \[
        (P_i-P_{ij})(P_j-P_{ij}) = P_iP_j - P_{ij} - P_{ij} + P_{ij} = P_iP_j - P_{ij}.
    \]
   Since orthogonal projections are self-adjoint, we have
    \[
        \langle (P_i-P_{ij})f, (P_j-P_{ij})g \rangle = \langle f, (P_iP_j-P_{ij})g \rangle
    \]
    for all $f,g \in \mathcal{H}$. Since $P_i-P_{ij}$ and $P_j-P_{ij}$ are orthogonal projections onto the subspaces of $W_i$ and $W_j$ (respectively) which are orthogonal to $W_{ij}$, we thus have
    \[
        \cos(W_i,W_j) = \sup_{\substack{f,g \in \mathcal{H} \\ f,g \neq 0}} \frac{\langle (P_i-P_{ij})f, (P_j-P_{ij})g \rangle}{\|f\| \, \|g\|} = \sup_{\substack{f,g \in \mathcal{H} \\ f,g \neq 0}} \frac{\langle f, (P_iP_j-P_{ij})g \rangle}{\|f\| \, \|g\|} = \|P_iP_j-P_{ij}\|,
    \]
    as desired. Note for the first equality that we can put $\|f\| \, \|g\|$ in the denominator instead of $\|(P_i-P_{ij})f\| \, \|(P_j-P_{ij})g\|$ because the $\sup$ will be maximized at $f$ and $g$ which are in the range of $P_i-P_{ij}$ and $P_j-P_{ij}$ respectively.
\end{proof}

Another natural property of the angle (or cosine) of two subspaces is that it should be preserved under taking orthogonal complements. Consider the simplest case, of two linear subspaces $L_u,L_w \in \R^2$. It is a basic geometry fact that the minimal angle between the lines $L_u$ and $L_w$ is the same as the minimal angle between the lines $L_u^\perp$ and $L_w^\perp$. This also holds for general subspaces.

\begin{lemma}[Orthogonal complements preserve angles {\cite{Kas11}}]
\label{lem:orth-complement}
For closed subspaces $U,W\subseteq\mathcal H$,
\[
    \cos(U,W)=\cos(U^\perp,W^\perp).
\]
\end{lemma}

\subsection{Equivalent forms of spectral gaps}\label{subsec:spectral-gaps}

We note that we do not actually use much of this section in the rest of the paper, but we keep it here to demonstrate the connections between the various spectral-gap-like quantities that appear here and in the surrounding literature.

Let $\mu$ be a probability measure on
$\Omega=\prod_{i\in V}\Omega_i$, where $V$ is finite and each
$\Omega_i$ is a standard Borel space of spins for site $i$, and let
$X=(X_i)_{i\in V}$ have law $\mu$. We will consider spectral gaps
for pinnings of all but $k$ sites in $V$. For $R\subseteq V$ with
$|R|=k$, let $\mu_\tau$ be a regular conditional distribution of $X_R$
given $X_{V\setminus R}=\tau$. This corresponds to a codimension-$k$
link. Note that 
all statements about pinnings are implicitly taken to mean to hold a.e. with respect to the law of the pinned spins.

In this section we give several equivalent interpretations of the
spectral gaps of $k$-links. We begin by defining several operators that will be featured in this equivalence.

\paragraph{Glauber dynamics on a link.}
Fix $\nu=\mu_\tau$ on the spins in $R$. Define
\[
    P_i^\nu f=\mathbb E_\nu[f\mid X_{R\setminus\{i\}}] \quad\text{and}\quad L_i^\nu=I-P_i^\nu.
\]
Both are orthogonal projections on $L^2(\nu)$.Since Glauber dynamics
chooses a site uniformly and resamples its spin conditioned on the
other spins, we have the corresponding operators
\[
    P_{\mathrm{Glaub},\nu}=\frac1k\sum_{i\in R}P_i^\nu \quad\text{and}\quad
    L_{\mathrm{Glaub},\nu}=I-P_{\mathrm{Glaub},\nu}
                         =\frac1k\sum_{i\in R}L_i^\nu.
\]
The link is \emph{connected} if
$\ker L_{\mathrm{Glaub},\nu}=\bigcap_{i\in R}\ker L_i^\nu$
consists only of constant functions.

Writing $L^2_0(\nu)$ for the subspace of functions that have zero mean w.r.t. $\nu$, we define
\[
\begin{aligned}
    \lambda_2(P_{\mathrm{Glaub},\nu})
       &:=\sup\sigma\!\left(
             P_{\mathrm{Glaub},\nu}|_{L^2_0(\nu)}\right),\\
    \lambda_2(L_{\mathrm{Glaub},\nu})
       &:=\inf\sigma\!\left(
             L_{\mathrm{Glaub},\nu}|_{L^2_0(\nu)}\right)
        =1-\lambda_2(P_{\mathrm{Glaub},\nu}).
\end{aligned}
\]
Because we are working over an infinite spin space, these spectral values may not actually be eigenvalues. 

\paragraph{Maximal correlation.}
Given a probability measure $\nu$ on two variables $X,Y$ (e.g., $\mu$
conditioned on fixed spins for all but two sites), we define the
\textbf{Hirschfeld-Gebelein-R\'enyi (HGR) maximal correlation} to be
\begin{equation}\label{eq:HGR}
    \rho_\nu(X,Y)
       =\sup_{f,g}
          \frac{|\operatorname{Cov}_\nu(f(X),g(Y))|}
               {\sqrt{\operatorname{Var}_\nu(f(X))
                      \operatorname{Var}_\nu(g(Y))}},
\end{equation}
where the supremum is over square-integrable $f\in L^2(\nu_X)$ and $g\in L^2(\nu_Y)$ with positive variance, and $\nu_X,\nu_Y$ are the marginal laws. If this set of pairs is empty, then define $\rho_\nu(X,Y)$ to be $0$.

Let $U_i^\nu$ be the subspace of
mean-zero functions in $L^2(\nu)$ which can be written as a function
of the spins of all sites in $R\setminus\{i\}$ ($\nu$-a.e.), i.e. for $i \in R$ we have 
\[
    U_i^\nu
       =\operatorname{ran}P_i^\nu\cap L^2_0(\nu).
\]

We now extend the HGR maximal correlation to the $k$-link using the following
multivariate generalization, previously studied in the statistics
literature \cite{Koy87,GZ22}:
\begin{equation}\label{eq:link-generalized-correlation}
    c_\nu
       :=\sup_{\substack{
             g_i\in U_i^\nu\ (i\in R)\\
             \sum_{i\in R}\|g_i\|_2^2>0}}
          \frac{\displaystyle
                   \sum_{\substack{i,j\in R\\i\ne j}}
                       \langle g_i,g_j\rangle_\nu}
               {\displaystyle
                   (k-1)\sum_{i\in R}\|g_i\|_2^2}.
\end{equation}

\paragraph{Connection to a walk on partial configurations.}\label{sec:random-walk}
We define a random walk as follows: start from a configuration on $R\setminus\{i\}$, we sample the spin at $i$ conditioned on the other spins, then choose
$j\in R\setminus\{i\}$ uniformly at random, and erase the spin at $j$; the corresponding transition operator $\mathcal S_{\nu,0}:\bigoplus_{i\in R}U_i^\nu
                          \longrightarrow\bigoplus_{i\in R}U_i^\nu$ is defined to be 
\[
    (\mathcal S_{\nu,0}g)_i
       =\frac1{k-1}\sum_{j\ne i}P_i^\nu g_j.
\]
For $g,h \in \bigoplus_{i\in R}U_i^\nu$, we define the inner product
\[
    \langle g,h\rangle_\oplus
       :=\sum_{i\in R}\langle g_i,h_i\rangle_\nu.
\]Since $g_i,h_i\in U_i^\nu$, we have
$P_i^\nu g_i=g_i$ and $P_i^\nu h_i=h_i$. Thus, using the
self-adjointness of $P_i^\nu$ and interchanging $i,j$, we obtain
\[
    \langle g,\mathcal S_{\nu,0}h\rangle_\oplus =\frac1{k-1}\sum_{i\ne j}
             \langle g_i,P_i^\nu h_j\rangle_\nu=\frac1{k-1}\sum_{i\ne j}
             \langle g_i,h_j\rangle_\nu=\frac1{k-1}\sum_{i\ne j}
             \langle P_i^\nu g_j,h_i\rangle_\nu=\langle\mathcal S_{\nu,0}g,h\rangle_\oplus,
\]
showing that $\mathcal S_{\nu,0}$ is self-adjoint. Furthermore, we have
\begin{equation}\label{eq:link-walk-correlation}
    \sup\sigma(\mathcal S_{\nu,0})=c_\nu.
\end{equation}

For two sites, this is the bipartite walk that alternates between
sampling $Y$ conditioned on $X$ and sampling $X$ conditioned on $Y$.
For any nontrivial two-variable law $\nu$, we define $T_\nu:L^2_0(\nu_Y)\longrightarrow L^2_0(\nu_X)$ by $ (T_\nu g)(x)=\mathbb E_\nu[g(Y)\mid X=x]$. We have
$\rho_\nu(X,Y)=\|T_\nu\|_{\mathrm{op}}$.
For $g \in L^2_0(\nu_X)$ and $h \in L^2_0(\nu_Y)$, we have \[
    \begin{pmatrix}g\\h\end{pmatrix}
    \longmapsto
    \begin{pmatrix}
        \mathbb E_\nu[h(Y)\mid X]\\
        \mathbb E_\nu[g(X)\mid Y]
    \end{pmatrix}
    =
    \begin{pmatrix}0&T_\nu\\T_\nu^*&0\end{pmatrix}
    \begin{pmatrix}g\\h\end{pmatrix}.
\]

Its spectrum is symmetric about zero, so its largest spectral value is
$\|T_\nu\|_{\mathrm{op}}=\rho_\nu(X,Y)$. That is,
$c_\nu=\rho_\nu(X,Y)$ when $k=2$.

\paragraph{Conditional-expectation and Laplacian angles.}
Finally, on 
$\bigoplus_{i\in R}L^2_0(\nu)$ we also define
\[
    \mathbf P_\nu=\operatorname{diag}(P_i^\nu)_{i\in R},
    \quad\text{and}\quad
    \mathbf L_\nu=\operatorname{diag}(L_i^\nu)_{i\in R}
                 =I-\mathbf P_\nu.
\]
Let $\Pi_{\Delta_\nu}$ be the orthogonal projection onto the  subspace $\Delta_\nu
       =\{(f,\ldots,f):f\in L^2_0(\nu)\}$, i.e.
\[
    \Pi_{\Delta_\nu}(g_i)_{i\in R}
       =\left(\frac1k\sum_i g_i,\ldots,
               \frac1k\sum_i g_i\right).
\]
We let $\Pi_{\mathcal N_\nu}$ denote the orthogonal projection onto
\[
    \mathcal N_\nu
       =\left\{(g_i)_{i\in R}:
           g_i\in\operatorname{ran}L_i^\nu,\quad
           \sum_i g_i=0\right\}.
\]
For $i\ne j$, we also let $P_{ij}^\nu$
and $L_{ij}^\nu$ be the orthogonal projections onto
$\operatorname{ran}P_i^\nu\cap\operatorname{ran}P_j^\nu$ and
$\operatorname{ran}L_i^\nu\cap\operatorname{ran}L_j^\nu$,
respectively.

We are now ready to state the equivalent forms of the spectral gap
on a $k$-link.

\begin{lemma}[Equivalent forms of link spectral gaps] \label{lem:equiv-forms}
Let $\nu=\mu_\tau$ be a connected, nontrivial codimension-$k$ link,
where $k\ge2$. For $0\le\gamma<1$, the following conditions are
equivalent:
\begin{enumerate}
    \item \textbf{Glauber dynamics spectral gap:}
    $\lambda_2(L_{\mathrm{Glaub},\nu})\ge(1-\gamma)/k$,
    or equivalently
    $\lambda_2(P_{\mathrm{Glaub},\nu})
        \le1-(1-\gamma)/k$;

    \item \textbf{Generalized correlation:}
    \[
        c_\nu\le\frac{k-2+\gamma}{k-1};
    \]

    \item \textbf{Partial-configuration walk:}
    \[
        \sup\sigma(\mathcal S_{\nu,0})
            \le\frac{k-2+\gamma}{k-1};
    \]

    \item \textbf{Conditional-expectation:}
    \[
        \|\mathbf P_\nu\Pi_{\Delta_\nu}\|_{\mathrm{op}}^2
            \le1-\frac{1-\gamma}{k};
    \]

    \item \textbf{Laplacian:}
    \[
        \|\mathbf L_\nu(I-\Pi_{\Delta_\nu})
            -\Pi_{\mathcal N_\nu}\|_{\mathrm{op}}^2
            \le1-\frac{1-\gamma}{k}.
    \]
\end{enumerate}
\end{lemma}

As an example, when $k=2$ say $R=\{i,j\}$, then we have the
HGR maximal correlation in \eqref{eq:HGR}, and
\[
    c_\nu=\rho_\nu(X_i,X_j)
       =\|P_i^\nu P_j^\nu-P_{ij}^\nu\|_{\mathrm{op}}
       =\|L_i^\nu L_j^\nu-L_{ij}^\nu\|_{\mathrm{op}}.
\]
\begin{proof}
Define $A:L^2_0(\nu)\longrightarrow\bigoplus_{i\in R}U_i^\nu$ by $Af=(P_i^\nu f)_{i\in R}$, so that  $A^*(g_i)=\sum_i g_i$. With this, we have 
\[
    A^*A=kP_{\mathrm{Glaub},\nu},
    \quad\text{and}\quad
    AA^*=I+(k-1)\mathcal S_{\nu,0}.
\]
Since $A^*A$ and $AA^*$ are nonnegative and have the same operator norm,
\eqref{eq:link-walk-correlation} implies that
\[
    k\lambda_2(P_{\mathrm{Glaub},\nu})=1+(k-1)c_\nu \quad\text{and}\quad
    \lambda_2(L_{\mathrm{Glaub},\nu})
       =\frac{k-1}{k}(1-c_\nu).
\]
This proves the equivalence of the first three conditions.

To relate these to the other two properties, we define the isometry $J:L^2_0(\nu)\longrightarrow\bigoplus_{i\in R}L^2_0(\nu)$ by  $Jf=\frac1{\sqrt{k}}(f,\ldots,f)$. Then $ \Pi_{\Delta_\nu}=JJ^*$ and for $f \in L^2_0(\nu)$ we have 
\[
    (J^*\mathbf P_\nu J)f
       =\frac1k\sum_iP_i^\nu f=P_{\mathrm{Glaub},\nu}f.
\]
Having this, we write
\[
    \|\mathbf P_\nu  \Pi_{\Delta_\nu}\|_{\mathrm{op}}^2
       =\| \Pi_{\Delta_\nu}\mathbf P_\nu  \Pi_{\Delta_\nu}\|_{\mathrm{op}}
       =\|J^*\mathbf P_\nu J\|_{\mathrm{op}}
       =\lambda_2(P_{\mathrm{Glaub},\nu}).
\]

Note that the preceding computations did not use the connectedness
of the link. That condition is precisely that $\bigcap_{i\in R}\operatorname{ran}P_i^\nu
       =\operatorname{span}\{1\}$.
Having this, we claim that
$\operatorname{ran}\mathbf P_\nu\cap\operatorname{ran} \Pi_{\Delta_\nu}=\{0\}$.
Indeed, a vector in $\operatorname{ran} \Pi_{\Delta_\nu}$ has the form
$(f,\ldots,f)$ with $f\in L^2_0(\nu)$. If it also belongs to
$\operatorname{ran}\mathbf P_\nu$, then we have for every $i \in R$ that $P_i^\nu f=f$ which implies $f$ is constant. Since $f$ has zero mean we in fact have $f \equiv 0$ as desired.

Now, since 
$\mathcal N_\nu
 =\operatorname{ran}\mathbf L_\nu\cap\operatorname{ran}(I- \Pi_{\Delta_\nu})$,
we have

\begin{align*}
    \|\mathbf P_\nu  \Pi_{\Delta_\nu}\|_{\mathrm{op}}&\underset{\text{Lemma~\ref{lem:angle-proj-norm}}}{=}
      \cos\bigl(\operatorname{ran}\mathbf P_\nu,
                \operatorname{ran} \Pi_{\Delta_\nu}\bigr)\underset{\text{Lemma~\ref{lem:orth-complement}}}{=}
      \cos\bigl((\operatorname{ran}\mathbf P_\nu)^\perp,
                (\operatorname{ran} \Pi_{\Delta_\nu})^\perp\bigr)\\
                &\underset{\mathbf L_\nu=I-\mathbf P_\nu}{=}
      \cos\bigl(\operatorname{ran}\mathbf L_\nu,
                \operatorname{ran}(I- \Pi_{\Delta_\nu})\bigr) \underset{\text{Lemma~\ref{lem:angle-proj-norm}}}{=}
      \|\mathbf L_\nu(I- \Pi_{\Delta_\nu})-\Pi_{\mathcal N_\nu}\|_{\mathrm{op}}.
\end{align*}
This proves the remaining equivalences and completes the proof.
\end{proof}

\paragraph{Connection to local spectral expansion.}

Here we show the connection back to local spectral expansion and spectral gaps for the base graphs of links of high-dimensional expanders. The following lemma is the crucial connection of the above notions which yields the proof of Corollary~\ref{cor:main-hdx-corr}.

\begin{lemma}[Multipartite comparison; see Proposition~30 of \cite{HR26}] \label{lem:multi-comp}
    Fix $k \geq 2$, and consider finite random variables $(Y_1,\ldots,Y_k)$ with joint distribution $\mu$. Let $W$ be the normalized adjacency operator of the weighted $k$-partite complex with this facet distribution. Then
    \[
        \lambda_2(W)\le \max_{a\ne b}\rho_\mu(Y_a,Y_b).
    \]
\end{lemma}

This follows from the multipartite comparison in Proposition~30 of \cite{HR26}, since the graph $K$ in the statement of their result is the complete graph here and has no positive non-trivial eigenvalues.

The proof of Corollary~\ref{cor:main-hdx-corr} then follows by bounding $\lambda_2(W_\tau)$ via the two-block conclusions of Theorem~\ref{thm:main-full}. After pinning $\tau$, the comparison matrix is $H_U$. Applying Theorem~\ref{thm:main-full} to each pair of remaining blocks, together with Schur-complement monotonicity from Lemma~\ref{lem:compression-schur}, gives a Glauber spectral gap of at least \((1-\lambda_{\max}(H_U))/2\) for every two-block marginal. Lemmas \ref{lem:equiv-forms} and \ref{lem:multi-comp} therefore imply
\[
    \lambda_2(W_\tau) \leq \lambda_{\max}(H_U) \leq \lambda_{\max}(H),
\]
as required.

\subsection{Vector calculus}
We write $\nabla$, $\operatorname{div}$, and $\Delta$ for the Euclidean gradient, divergence, and Laplacian, respectively. Let $\mathbb B^N$denote the open unit ball in $\mathbb R^N$, and let $\omega$ be the uniform probability measure on $\mathbb S^{N-1}$. We also let $\omega_N$ the volume of $\mathbb B^N$.

We next recall the divergence theorem. We will apply the divergence theorem to Lipschitz functions which are initially
defined only on the sphere. To that end, we first extend such functions to the
whole unit ball without increasing the Lipschitz constant.

\begin{proposition}[McShane-Whitney extension theorem]
\label{prop:whitney-extension}
Let $f:\mathbb S^{N-1}\to\mathbb R$ be $\ell$-Lipschitz with respect
to Euclidean distance. Then $f$ has an $\ell$-Lipschitz extension
to $\mathbb R^N$, given by
\[
    F(x)=\inf_{u\in\mathbb S^{N-1}}
             \bigl(f(u)+\ell\|x-u\|\bigr).
\]
In particular, $F|_{\mathbb S^{N-1}}=f$. Furthermore, Rademacher's theorem shows $F$ is differentiable a.e.
and $\|\nabla F\|\le\ell$ a.e.
\end{proposition}

Using this extension, we can state the divergence
theorem for Lipschitz functions on $\mathbb{R}^N$ and the form in which we will be most amenable for our application. 

\begin{proposition}[Divergence theorem]
\label{fact:divergence-lipschitz}
Let $F:\mathbb R^N\to\mathbb R$ be Lipschitz and let
$V\in C^1(\overline{\mathbb B^N};\mathbb R^N)$. Then
\[
    \int_{\mathbb S^{N-1}}F(u)V(u)\cdot u\,d\omega(u)
       =\frac1{N\omega_N}\int_{\mathbb B^N}
          \bigl(\nabla F\cdot V+F\operatorname{div}V\bigr)\,dx.
\]

Consequently, if $\operatorname{div}V=0$ in $\mathbb B^N$ and
$g(u)=V(u)\cdot u$ on $\mathbb S^{N-1}$, then for every
$\ell$-Lipschitz function $f:\mathbb S^{N-1}\to\mathbb R$, we have
\begin{equation}\label{eq:prelim-flux-bound}
    \left|\int_{\mathbb S^{N-1}}f g\,d\omega\right|
       \le\frac{\ell}{N\omega_N}
          \int_{\mathbb B^N}\|V(x)\|\,dx.
\end{equation}
\end{proposition}
\begin{proof}
For smooth $F$, we apply the divergence theorem to $FV$, and for Lipschitz $F$, we make smooth approximations and pass to the limit. For the final bound, it follows by combining
Proposition~\ref{prop:whitney-extension} and the fact that
$|\nabla F\cdot V|\le\ell\|V\|$ a.e.
\end{proof}
\subsection{Real stable polynomials}
A polynomial $f\in\mathbb{R}[z_1,\ldots,z_d]$ is called
\textbf{real stable} if $f(z_1,\ldots,z_d)\ne0$ whenever
$\operatorname{Im}z_i>0$ for every $i$. We also regard the zero
polynomial as real stable. In one variable, a nonzero polynomial is
real stable precisely when all its roots are real. A polynomial is
\textbf{multiaffine} if it has degree at most one in each variable.
Real stability is preserved by products, partial differentiation,
and specialization of variables to real constants
\cite[Section~6]{MSS15}.

We will use the following inequalities, usually called the
\textbf{Rayleigh inequalities} (see for example \cite{Bra07}).

\begin{fact}[Rayleigh inequalities]
\label{fact:rayleigh}
If $f\in\mathbb{R}[z_1,\ldots,z_d]$ is real stable, then
\[
    (\partial_i f)(x)(\partial_j f)(x)
        \ge f(x)(\partial_i\partial_j f)(x)
    \qquad (x\in\mathbb{R}^d,\ i\ne j).
\]
At points where $f(x)>0$, this is equivalently
$\partial_i\partial_j\log f(x)\le0$.
For multiaffine polynomials, the Rayleigh inequalities also characterize
real stability.
\end{fact}

We also record the following consequence of the Lieb-Sokal differential rule which tells about one operation that preserves real stability (see for example \cite{MSS15}).  

\begin{fact}
\label{fact:stable-pair-derivative}
Suppose that $f$ is real stable and multiaffine in the variables
$x_u,x_v$, where $u\ne v$. Then, for every $c\ge0$, we have that $(1-c\,\partial_{x_u}\partial_{x_v})f
$ is real stable.
\end{fact}

We will also use the following log-concavity consequence of real stability for
polynomials with nonnegative coefficients (for a proof see for example \cite{AOV21}).

\begin{fact}[Log-concavity]
\label{fact:stable-log-concave}
Let $f\in\mathbb R[z_1,\ldots,z_d]$ be a nonzero real stable
polynomial with nonnegative coefficients. Then $\log f$ is concave
on $\mathbb R_{>0}^d$, or equivalently, for $x\in\mathbb R_{>0}^d$, we have  $ \nabla^2\log f(x)
      \preceq0$.
By continuity, the same inequality holds at every
$x\in\mathbb R_{\ge0}^d$ for which $f(x)>0$.
\end{fact}

\section{Orthogonal Projections and the Proofs of Theorems \ref{thm:main-prev-Bochner} and \ref{thm:main-full}} \label{sec:orthog-proj}

In this section, we prove our main technical results, which give local-to-global type results for orthogonal projections on Hilbert spaces. These general results then imply our main result Theorem~\ref{thm:main-full} and also Theorem~\ref{thm:main-prev-Bochner} along the way. We start with a warm-up result in Section~\ref{sec:Hilbert-small-result}, and then state and prove the main technical result Theorem~\ref{thm:abstract-proj}. Finally in Section~\ref{sec:proof-of-main-full} we use this to prove Theorem~\ref{thm:main-full}.

Before moving on, we note that some of the bounds developed below can be obtained from Kassabov’s results on angles between subspaces (Theorem~5.1 and Section~5.2 of \cite{Kas11}), applied to the kernels of the orthogonal projections we study. In particular, his subspace-angle estimates imply the global Poincar\'e bound and, through successive intersections, the projection norm bounds in Theorem~\ref{thm:abstract-proj} for singleton sets $T_i$. His successive-intersection estimates correspond to taking and diagonally normalizing Schur complements of the matrix $M$ in Theorem~\ref{thm:abstract-proj} and Theorem~\ref{thm:main-full}. We give a self-contained proof using Bochner operators and their compressions that treats arbitrary block partitions. This allows for a direct treatment of these inequalities in the language of spin systems and makes their dependence on the matrix of conditional pairwise bounds explicit.

\subsection{A warm-up result} \label{sec:Hilbert-small-result}

We first prove very general bounds on the Bochner operator discussed in Section~\ref{sec:tech-overview} above, along with a few nice identities which we will make use of in proving the full result in Theorem~\ref{thm:abstract-proj}.

\begin{lemma}[Bochner operator lemma] \label{lem:Bochner-lite}
    Let $(\mathcal{H}_i, \langle \cdot, \cdot \rangle_i)$ be Hilbert spaces for $i \in \{0,1,2\}$, and let $d_0: \mathcal{H}_0 \to \mathcal{H}_1$ and $d_1: \mathcal{H}_1 \to \mathcal{H}_2$ be bounded linear maps such that $d_1d_0 = 0$. Define $Q: \mathcal{H}_0 \to \mathcal{H}_0$ and $B: \mathcal{H}_1 \to \mathcal{H}_1$ via
    \[
        Q = d_0^*d_0 \quad \text{and} \quad B = d_0d_0^* + d_1^*d_1.
    \]
    If $B\succeq\alpha I$ for some $\alpha>0$, then $B$ is
boundedly invertible and
\[
    Q^2=d_0^*Bd_0\succeq\alpha Q\quad\text{and}\quad P=d_0^*B^{-1}d_0\preceq\alpha^{-1}Q.
\]
    where $P$ is the orthogonal projection onto $(\ker d_0)^\perp$.
\end{lemma}
\begin{proof}
    Since $B$ is self-adjoint, the fact that $B \geq \alpha I$ for $\alpha > 0$ implies $B^{-1} \leq \alpha^{-1} I$ is a standard fact (even in infinite dimensions).
    
    We first compute some operator identities. We have
    \[
        Bd_0 = d_0d_0^*d_0 + d_1^*d_1d_0 = d_0 Q,
    \]
    which implies $Q^2 = d_0^* B d_0$. It also implies
    \[
        d_0^* B^{-1} d_0 Q = d_0^* B^{-1} B d_0 = d_0^* d_0 = Q.
    \]
    Thus $d_0^* B^{-1} d_0$ acts as the identity on $\operatorname{ran}(Q)$, and it is clear that $d_0^*B^{-1}d_0$ vanishes on $\ker d_0$. Since $Q$ is self-adjoint, we have $(\ker d_0)^\perp = (\ker Q)^\perp = \overline{\operatorname{ran}(Q)}$. Therefore $P = d_0^*B^{-1}d_0$ since $d_0^* B^{-1} d_0$ is bounded.

    With these identities, we now prove the desired operator inequalities. We have
    \[
        Q^2 = d_0^* B d_0 \succeq \alpha d_0^* d_0 = \alpha Q \quad\text{and}\quad
        P = d_0^*B^{-1}d_0 \preceq \alpha^{-1} d_0^* d_0 = \alpha^{-1} Q
    \]
    as desired.
\end{proof}

\subsection{General result}

We now prove our main technical result on Hilbert space projections, which leads to our main result Theorem~\ref{thm:main-full}. We first need a bit of setup.

Let $\mathcal{H}$ be a (possibly infinite dimensional) real Hilbert space. Let $W_1,\ldots,W_n \subseteq \mathcal{H}$ be closed subspaces, and let $P_1,\ldots,P_n$ be the corresponding orthogonal projections onto $W_1,\ldots,W_n$ respectively. Further define $P_{ij}$ to be projections onto $W_{ij} = W_i \cap W_j$ for all $i \neq j$. Finally, let $W = \overline{\sum_{i=1}^n W_i}$, and let $P$ be the projection onto $W$.

Additionally, we fix disjoint nonempty subsets $T_1,\ldots,T_m \subset [n]$ and let $T = T_1 \sqcup \cdots \sqcup T_m$. We consider a restriction of the above picture as follows. Let $\mathcal{H}^T = \bigcap_{i \not\in T} W_i^\perp$ and
\[
    \tilde{W}_i = \mathcal{H}^T \cap (\mathcal{H}^T \cap W_i^\perp)^\perp,
\]
and define $\tilde{P}_i: \mathcal{H}^T \to \mathcal{H}^T$ to be the orthogonal projection onto $\tilde{W}_i$ for all $i \in [n]$. We now further define $\tilde{Q}_a: \mathcal{H}^T \to \mathcal{H}^T$ to be the orthogonal projection onto $\overline{\sum_{i \in T_a} \tilde{W}_i}$ for all $a \in [m]$, and we define $\tilde{P}: \mathcal{H}^T \to \mathcal{H}^T$ to be the orthogonal projection onto $\overline{\sum_{i \in T} \tilde{W}_i}$.

\begin{theorem} \label{thm:abstract-proj}
    Let notation be as above, and suppose that $\|P_iP_j - P_{ij}\| \leq s_{ij}$ for all $i \neq j$ for a given real symmetric matrix $S$ with non-negative entries, zero diagonal, and maximum eigenvalue strictly less than $1$. Further define
    \[
        M = I-S, \quad C = ((M^{-1})_T)^{-1}, \quad D = \operatorname{blockdiag}(C_{T_1}, \ldots, C_{T_m})
    \]
    and
    \[
        \alpha = \lambda_{\min}(D^{-1/2}CD^{-1/2}).
    \]
    Then
    \[
        \|\tilde{P}f\|^2 \leq \alpha^{-1} \sum_{a=1}^m \|\tilde{Q}_a f\|^2
    \]
    for all $f \in \mathcal{H}^T$.
\end{theorem}

The assumption of $\|P_iP_j - P_{ij}\| \leq s_{ij}$ is saying something about how close $P_i$ and $P_j$ are to commuting, or equivalently, how close $W_i$ and $W_j$ are to being orthogonal once you remove their intersection. This is a bound on how small the angle between two elements of $W_i$ and $W_j$, which are orthogonal to $W_i \cap W_j$, can be. This and other equivalent formulations of this condition were discussed in Section~\ref{subsec:spectral-gaps}.

\subsection{Proof of Theorem ~\ref{thm:abstract-proj}} \label{sec:proof-abstract-proj}

\paragraph{Hilbert spaces and maps between them.}
Define
\[
    \mathcal{H}_0 = \mathcal{H}, \quad \mathcal{H}_1 = \bigoplus_i W_i, \quad \mathcal{H}_2 = \bigoplus_{i<j} W_{ij}
\]
with inner products given by
\[
    \langle f, g \rangle_0 = \langle f, g \rangle_{\mathcal{H}}, \quad \langle (f_i)_i, (g_i)_i \rangle_1 = \sum_i \langle f_i, g_i \rangle_{\mathcal{H}}, \quad \langle (f_{ij})_{i<j}, (g_{ij})_{i<j} \rangle_2 = \sum_{i<j} \langle f_{ij}, g_{ij} \rangle_{\mathcal{H}}.
\]
Further define
\[
    d_0: \mathcal{H}_0 \to \mathcal{H}_1 \quad \text{via} \quad d_0(f)_i = P_i f \quad \text{for all $i$}
\]
and
\[
    d_1: \mathcal{H}_1 \to \mathcal{H}_2 \quad \text{via} \quad d_1\big((f_i)_i\big)_{ij} = P_{ij}(f_i-f_j) \quad \text{for all $i<j$}.
\]
These maps are both bounded. Note that $d_1d_0 = 0$, which follows from $P_{ij}P_i=P_{ij}P_j=P_{ij}$. We also have
\begin{itemize}
    \item $d_0^*\big((f_i)_i\big) = \sum_i f_i$,
    \item $d_1^*\big((f_{ij})_{i,j}\big)_i = \sum_{j > i} f_{ij} - \sum_{j < i} f_{ji}$,
    \item $d_0^*d_0 = \sum_i P_i$, and
    \item $(\ker d_0)^\perp = W$.
\end{itemize}

\paragraph{Bochner operator.}
Now define $B: \mathcal{H}_1 \to \mathcal{H}_1$ via $B = d_0d_0^* + d_1^*d_1$. Therefore,
\[
    B(f)_i = P_i\left(\sum_j f_j\right) + \sum_{j \neq i} P_{ij}(f_i-f_j),
\]
or equivalently on blocks $B_{ij}: W_j \to W_i$ via
\[
    B_{ij} = \begin{cases}
        I + \sum_{k \neq i} P_{ik}, & i = j \\
        P_iP_j - P_{ij}, & i \neq j
    \end{cases}.
\]
\paragraph{Coercivity of the Bochner operator.}
From our expression for $B$, we have $B_{ii} \succeq I$ and $\|B_{ij}\| \leq s_{ij}$ for $i \neq j$ by assumption. Given $(f_i)_i \in \mathcal{H}_1$, we write $a_i = \|f_i\|$ for all $i$. This implies
\[
    \langle f, Bf \rangle_1 = \sum_i \langle f_i, B_{ii} f_i \rangle_0 + \sum_{i \neq j} \langle f_i, B_{ij} f_j\rangle_0 
        \geq \sum_i a_i^2 - \sum_{i \neq j} s_{ij} a_i a_j 
        = a^\top (I-S) a 
        \geq \lambda_{\min}(M) \sum_i a_i^2,
\]
and thus $B \succeq \lambda_{\min}(M) I$. Therefore, $B$ is boundedly invertible and $B^{-1} \preceq \lambda_{\min}(M)^{-1} I$.

Note also by Lemma~\ref{lem:Bochner-lite} and $(\ker d_0)^\perp = W$ we have $P = d_0^* B^{-1} d_0$. Applying Lemma~\ref{lem:Bochner-lite} with $\alpha = \lambda_{\min}(M) \geq \epsilon$ gives both inequalities in Proposition~\ref{prop:1-2-tech-overview}. This implies Theorem~\ref{thm:main-prev-Bochner}, by the same specialization to spin systems as in Section~\ref{sec:proof-of-main-full}.

\paragraph{Restricting to $T$.}
Let $R = [n] \setminus T$. Define
\[
    \mathcal{H}_0^T = \mathcal{H}^T = \bigcap_{i\notin T}W_i^\perp,\qquad
    \mathcal{H}_1^T=\bigoplus_{i\in T}W_i,
\]
and define
\[
    d_0^T: \mathcal{H}_0^T \to \mathcal{H}^1_T, \qquad d_0^T f = (P_if)_{i \in T}.
\]
We use the decomposition $\mathcal{H}_1 = \mathcal{H}_1^T \oplus \bigoplus_{i \in R} W_i$. The compression lemma guarantees the inverses below exist.

Now define $\tilde{B}: \mathcal{H}^T_1 \to \mathcal{H}^T_1$ via
\[
    \tilde{B} = ((B^{-1})_{TT})^{-1} = B_{TT} - B_{TR} B_{RR}^{-1} B_{RT}.
\]
Similarly, note that
\[
    C = M_{TT} - M_{TR} M_{RR}^{-1} M_{RT}.
\]
When $R = \varnothing$, we omit the subtracted terms.

Note also that Lemma~\ref{lem:compression-schur} gives
\[
    \lambda_{\min}(M) I \preceq \tilde{B} \preceq \|B\| I
\]
Applying Lemma~\ref{lem:compression-schur} to $M$, we also have that $C$ and the principal blocks of $C$ are positive definite. Thus $D$ is also positive definite.

\paragraph{Outer block bound on $B^{-1}$.}
Here we prove that
\begin{equation} \label{eq:outer-block-bound}
    \|(B_{RR}^{-1})_{k\ell}\| \leq (M_{RR}^{-1})_{k\ell}
\end{equation}
for all blocks $k,\ell \in R$. To see this, fix $y,z$ such that $B_{RR} y = z$. Thus $z_k = B_{kk} y_k + \sum_{\ell \in R, \ell \neq k} B_{k\ell} y_\ell$, and since $B_{kk} \succeq I$ we have
\[
    \|z_k\| \geq \|B_{kk} y_k\| - \sum_{\ell \in R, \ell \neq k} \|B_{k\ell} y_\ell\| \geq \|y_k\| - \sum_{\ell \in R, \ell \neq k} s_{k\ell} \|y_\ell\|.
\]
That is, letting $a_i = \|y_i\|$ and $b_i = \|z_i\|$ for all $i \in R$ we have
\[
    (I-S_{RR})a \leq b
\]
entry-wise. Since $M_{RR}^{-1} = (I-S_{RR})^{-1} = \sum_{t=0}^\infty S_{RR}^t$ has non-negative entries, we obtain
\[
    a \leq M_{RR}^{-1} b
\]
entry-wise.

Fix $\ell\in R$ and choose any $z_\ell\in W_\ell$. Set $z_t=0$ for $t \neq \ell$, and let $y = B_{RR}^{-1}z$. Applying the above inequality yields
\[
    \|(B_{RR}^{-1})_{k\ell} z_\ell\| = \|(B_{RR}^{-1} z)_k\| = a_k \leq (M_{RR}^{-1} b)_k = (M_{RR}^{-1})_{k\ell} \|z_\ell\|,
\]
which implies $\|(B_{RR}^{-1})_{k\ell}\| \leq (M_{RR}^{-1})_{k\ell}$.

\paragraph{Block-wise bounds on $\tilde{B}$.}
Now for all $i,j \in T$ we have
\[
    \|(B_{TR} B_{RR}^{-1} B_{RT})_{ij}\| \leq \sum_{k,\ell \in R} \|B_{ik}\| \|(B_{RR}^{-1})_{k\ell}\| \|B_{\ell j}\| \underset{\eqref{eq:outer-block-bound}}{\leq} \sum_{k,\ell \in R} s_{ik} (M_{RR}^{-1})_{k\ell} s_{\ell j} = (S_{TR} M_{RR}^{-1} S_{RT})_{ij}.
\]
Thus for all $i \in T$, since $B_{ii} \succeq I$ we have
\[
    \tilde{B}_{ii} = B_{ii} - B_{iR}B_{RR}^{-1}B_{Ri} \succeq \big(1 - (S_{TR} M_{RR}^{-1} S_{RT})_{ii}\big) I.
\]
And for all $i \neq j$ in $T$ we have
\[
    \|\tilde{B}_{ij}\| = \|B_{ij} - B_{iR}B_{RR}^{-1}B_{Rj}\| \leq \|B_{ij}\| + \|(B_{TR} B_{RR}^{-1} B_{RT})_{ij}\| \leq s_{ij} + (S_{TR} M_{RR}^{-1} S_{RT})_{ij}.
\]
Since $T,R$ are disjoint, $S_{RT} = -M_{RT}$ and $S_{TR} = -M_{TR}$. Therefore, for $i \in T$ we have
\begin{equation} \label{eq:B-tilde-diag}
    \tilde{B}_{ii} \succeq (1 - (M_{TR} M_{RR}^{-1} M_{RT})_{ii})I = c_{ii} I
\end{equation}
and for all $i \neq j$ in $T$ we have
\begin{equation} \label{eq:B-tilde-off-diag}
    \|\tilde{B}_{ij}\| \leq -m_{ij} + (M_{TR} M_{RR}^{-1} M_{RT})_{ij} = -c_{ij}.
\end{equation}
Note that these inequalities imply for all $f = (f_i)_{i \in T} \in \mathcal{H}_1^T$ and all $i,j \in T$ that
\begin{equation} \label{eq:B-tilde-norm-ineq}
    \langle f_i, \tilde{B}_{ij} f_j \rangle_0 \geq c_{ij} \|f_i\| \, \|f_j\|.
\end{equation}

\paragraph{Coercivity-type inequality for $\tilde{B}$.}
Now, we define $\tilde{D} = \operatorname{blockdiag}(\tilde{B}_{T_1T_1}, \ldots, \tilde{B}_{T_mT_m})$. By Lemma~\ref{lem:compression-schur}, each diagonal block satisfies
\[
    \tilde{B}_{T_aT_a} \succeq \lambda_{\min}(M)I.
\]
Thus, it follows that $\tilde{D}$ is boundedly invertible.

Here we prove
\begin{equation} \label{eq:B-tilde-coercivity}
    \tilde{B} \succeq \alpha \tilde{D}.
\end{equation}
Note that by definition of $\alpha = \lambda_{\min}(D^{-1/2}CD^{-1/2})$, we have $C \succeq \alpha D$. Further, we have $0 < \alpha \leq 1$ by positive-definiteness of $C$ and the fact that $D^{-1/2}CD^{-1/2}$ has diagonal entries equal to $1$.

Now fix $f = (f_i)_{i \in T} \in \mathcal{H}_1^T$ and define $a_i = \|f_i\|$ for all $i$. By \eqref{eq:B-tilde-norm-ineq} we have
\begin{align*}
    \langle f, (\tilde{B}-\alpha\tilde{D}) f \rangle_1 &= (1-\alpha) \sum_a \sum_{i,j \in T_a} \langle f_i, \tilde{B}_{ij} f_j \rangle_0 + \sum_{a \neq b} \sum_{i \in T_a, j \in T_b} \langle f_i, \tilde{B}_{ij} f_j \rangle_0 \\
        &\geq (1-\alpha) \sum_a \sum_{i,j \in T_a} c_{ij} a_i a_j + \sum_{a \neq b} \sum_{i \in T_a, j \in T_b} c_{ij} a_i a_j \\
        &= a^\top (C - \alpha D) a \\
        &\geq 0
\end{align*}
Therefore, we have $\tilde{B} \succeq \alpha\tilde{D}$ as desired. Note also that this implies $\tilde{D}^{-1} \succeq \alpha \tilde{B}^{-1}$ by Lemma~\ref{lem:compression-schur}.

\paragraph{Back to the projections.}
The definitions give the following kernels on $\mathcal{H}_0^T$:
\[
    \ker\tilde{P}=W^\perp \quad \text{and} \quad
    \ker\tilde{Q}_a = \mathcal{H}_0^T \cap \bigcap_{i\in T_a} W_i^\perp.
\]
Since $W^\perp\subseteq \mathcal{H}_0^T$, it follows that
\[
\operatorname{ran} \tilde{P} = \mathcal{H}_0^T \cap W \quad \text{and} \quad
\tilde{P}f=Pf, ~ \forall f \in \mathcal{H}_0^T.
\]
We also have $\operatorname{ran} \tilde{Q}_a \subseteq \operatorname{ran}\tilde{P}$.

Now fix $f \in \mathcal{H}_0^T$ and define $g_i = P_i f$ for all $i \in T$. Thus $g = d_0^T f$. Since $P_i f = 0$ for all $i \not\in T$, we have
\begin{equation} \label{eq:B-tilde-to-proj}
    \|\tilde{P}f\|^2 = \|Pf\|^2 = \langle d_0 f, B^{-1} d_0 f \rangle_1 = \langle g, (B^{-1})_{TT} g \rangle_1 = \langle g, \tilde{B}^{-1} g \rangle_1,
\end{equation}
where here we used $P = d_0^* B^{-1} d_0$ from Lemma~\ref{lem:Bochner-lite} and $d_0f=(g,0)$ in the decomposition above.

Now for each $a \in [m]$ let $h_a = \tilde{Q}_a f$ and $g^{(a)} = d_0^T h_a = (P_i h_a)_{i \in T}$. Since $h_a\in\operatorname{ran}\tilde{P}$, we have $Ph_a=h_a$. Moreover, 
\[
    f - h_a \in \ker\tilde{Q}_a = \mathcal{H}_0^T \cap \bigcap_{i\in T_a} W_i^\perp,
\]
and thus $P_i(f-h_a)=0$ for every $i\in T_a$. Therefore $g^{(a)}_{T_a}=g_{T_a}$.

Thus by applying \eqref{eq:B-tilde-to-proj} to $h_a$ and then applying part 3 of Lemma~\ref{lem:compression-schur} we have
\begin{equation} \label{eq:Q-tilde-a}
    \|\tilde{Q}_a f\|^2 = \langle g^{(a)}, \tilde{B}^{-1} g^{(a)} \rangle_1 \geq \langle g_{T_a}, (\tilde{B}_{T_aT_a})^{-1} g_{T_a} \rangle_1,
\end{equation}
where the inequality is part 3 of Lemma~\ref{lem:compression-schur} applied to $\tilde{B}$ and the coordinate subspace $\bigoplus_{i\in T_a}W_i\subseteq \mathcal{H}_1^T$.

We now combine everything above to obtain
\[
    \sum_{a=1}^m \|\tilde{Q}_a f\|^2 \underset{\eqref{eq:Q-tilde-a}}{\geq} \sum_{a=1}^m \langle g_{T_a}, (\tilde{B}_{T_aT_a})^{-1} g_{T_a} \rangle = \langle g, \tilde{D}^{-1} g \rangle \underset{\eqref{eq:B-tilde-coercivity}}{\geq} \alpha \langle g, \tilde{B}^{-1} g \rangle \underset{\eqref{eq:B-tilde-to-proj}}{=} \alpha \|\tilde{P}f\|^2,
\]
which is precisely the desired conclusion.

\subsection{Proof of Theorem~\ref{thm:main-full}} \label{sec:proof-of-main-full}

In this section, we put together the tools developed above to prove
our main result, Theorem~\ref{thm:main-full}.

\begin{proof}[Proof of Theorem~\ref{thm:main-full}]
First, note that interpreting the local assumption on links using Lemma~\ref{lem:equiv-forms} and averaging, we have 
\[
    \|L_iL_j-L_{ij}\|_{\mathrm{op}}\le a_{ij},
\]
This shows that the hypotheses of Theorem~\ref{thm:abstract-proj} hold for $\mathcal H=L^2(\mu)$,
$W_i=\operatorname{ran}L_i$, and $S$ s.t. $S_{ij} = a_{ij}$.

Now let $f=f(X_T)\in L^2(\mu)$. Note that
$f\in\mathcal H_T$, since $L_if=0$ for every $i\notin T$. By the assumed connectedness, 
\[
    \ker\tilde P
       =\bigcap_{i\in V}\ker L_i
         =\operatorname{span}\{1\} \quad\text{and}\quad
    \ker\tilde Q_a
       =\mathcal H_T\cap\bigcap_{i\in T_a}\ker L_i.
\]
Consequently, $\tilde Pf=f-\mathbb E_\mu f$. 

Having this, we write
\begin{align*}
    \alpha\operatorname{Var}_\mu(f) =\alpha\|f-\mathbb E_\mu f\|_2^2 = \alpha \|\tilde Pf\|_2^2
    &\underset{\text{Theorem~\ref{thm:abstract-proj}}}{\le} \sum_{a=1}^m\|\tilde Q_af\|_2^2 =\sum_{a=1}^m
      \left\|\tilde Q_a
        \bigl(f-\mathbb E_\mu[f\mid X_{T\setminus T_a}]\bigr)
      \right\|_2^2 \\
      &\underset{\|\tilde{Q}_a\|\leq 1}{\le}\sum_{a=1}^m
      \left\|f-\mathbb E_\mu[f\mid X_{T\setminus T_a}]\right\|_2^2 \underset{\text{orthogonality}}{=}m\left\langle
        f,L_{\mathrm{Glaub}}^{T_1,\ldots,T_m}f
      \right\rangle_\mu,
\end{align*}
      where the first equality follows from the fact that every function of $X_{T\setminus T_a}$ belongs to
$\ker\tilde Q_a$.

Since $\alpha
    =\lambda_{\min}(D^{-1/2}CD^{-1/2})
    =1-\lambda_{\max}(H)$, by Fact~\ref{fact:spectral-gap-equivalences} we have 
\[
    \lambda_2\bigl(L_{\mathrm{Glaub}}^{T_1,\ldots,T_m}\bigr)
       \ge\frac{1-\lambda_{\max}(H)}{m},
\]
as desired.
\end{proof}

\section{Application: Mixing time of $O(N)$ model} \label{sec:full-ON-app}

In this section, we prove the following theorem about the spectral gap of Glauber dynamics in the $O(N)$ model.

\maingap*

\begin{remark}\label{rem:marginal-corr}
We can also use Theorem~\ref{thm:main-full} to bound correlations and
analyze block dynamics for the $O(N)$ model. To see this, we fix
the spins at all sites other than $x,y$. The main observation is that the maximal correlation between
the two remaining spins is at most $|J_{xy}|/N$, for every
choice of the fixed spins and external fields
(see Lemma~\ref{lem:two-spin}). Having this, we can
apply Theorem~\ref{thm:main-full} with $S=J_{\mathrm{abs}}/N$ whenever
$\theta:=\|S\|_{\mathrm{op}}<1$.

First, taking $T_1=\{x\}$ and $T_2=\{y\}$ corresponds to
integrating out all spins other than $x,y$. The theorem then
gives, for any square-integrable $f=f(\sigma_x)$ and
$g=g(\sigma_y)$,
\[
    |\operatorname{Cov}_\mu(f,g)|
    \leq
    \min\left\{\theta,
        \frac{\theta^{\operatorname{dist}_G(x,y)}}{1-\theta}\right\}
    \sqrt{\operatorname{Var}_\mu(f)\operatorname{Var}_\mu(g)}.
\]
In general, for any partition $V=T_1\sqcup\cdots\sqcup T_m$,
Theorem~\ref{thm:main-full} gives a spectral gap of at least
$(1-\lambda_{\max}(H))/m$ for the block dynamics of choosing a block
uniformly and resampling its spins conditioned on the remaining spins.
\end{remark}

\subsection{Proof of Theorem~\ref{thm:main-gap}}

\begin{lemma}\label{lem:two-spin}
Let $a,b\in\R^N$ and $\kappa\in\R$, and suppose
\[
\nu(du,dv)=\frac1Z
\exp\bigl(a\cdot u+b\cdot v+\kappa u\cdot v\bigr)
\,d\omega(u)\,d\omega(v).
\]
Then $\rho_\nu(U,V)\le\min\{1,|\kappa|/N\}$.
\end{lemma}

Assuming Lemma~\ref{lem:two-spin}, we prove Theorem~\ref{thm:main-gap}.
\begin{proof}[Proof of Theorem~\ref{thm:main-gap}]
If $\|J_{\mathrm{abs}}\|_{\mathrm{op}}\geq N$, the spectral-gap bound follows from the
nonnegativity of $L_{\mathrm{Glaub}}$. We therefore assume
$\epsilon:=1-\|J_{\mathrm{abs}}\|_{\mathrm{op}}/N>0$, where
$J_{\mathrm{abs}}=(|J_{xy}|)_{x,y\in V}$.

Fix distinct $x,y\in V$ and condition on
$\sigma_{V\setminus\{x,y\}}=\tau$. The law of
$(\sigma_x,\sigma_y)$ on this link is 
\[
    \propto \exp\left[
        J_{xy}\sigma_x\cdot\sigma_y
        +\left(h_x+\sum_{z\neq x,y}J_{xz}\tau_z\right)\cdot\sigma_x
        +\left(h_y+\sum_{z\neq x,y}J_{yz}\tau_z\right)\cdot\sigma_y
    \right]d\omega(\sigma_x)\,d\omega(\sigma_y).
\]
By Lemma~\ref{lem:two-spin}, its maximal correlation is at most
$|J_{xy}|/N$, uniformly in $\tau$. Since $\mu$ has a strictly positive
density with respect to the product measure, the spin system is connected. 

Consequently, using Lemma\ref{lem:equiv-forms} to verify the local hypothesis, we apply Theorem\ref{thm:main-full} with $A=J_{\mathrm{abs}}/N$ all singleton blocks to get 
\[
    \lambda_2(L_{\mathrm{Glaub}})
      \geq\frac{1-\rho(S)}n=\frac\epsilon n\quad\text{and}\quad
    \lambda_2(P_{\mathrm{Glaub}})\leq1-\frac\epsilon n.
\]
This proves the spectral-gap bound.

Finally, we prove the mixing time consequence. After $s$ updates, note that the
probability that some site has not been refreshed is at most $ne^{-s/n}$.
Conditioned on every site having been refreshed and writing $\lambda$ for
this conditional law, we have
\[
 \frac{d\lambda}{d\mu}
 =\frac{d\lambda}{d\mu_{0,h}}\frac{d\mu_{0,h}}{d\mu}
 \le\left(\prod_x e^{2\sum_y|J_{xy}|}\right)
       e^{\sum_{x,y}|J_{xy}|}
 =e^{3\sum_{x,y}|J_{xy}|}.
\]
Consequently, we have
\[
 \|\mathcal L(\sigma_{s+t})-\mu\|_{\mathrm{TV}}
 \le ne^{-s/n}
    +\frac12\exp\Big(\frac{3\sum_{x,y}|J_{xy}|}{2}
                       -\frac{\epsilon t}{n}\Big).
\]
It follows that the mixing time of Glauber dynamics is therefore
$O\Big(n\log n+\frac{n(\sum_{x,y}|J_{xy}|+1)}{\epsilon}\Big)
=O_{N,\epsilon}(n^2)$, since
$\sum_{x,y}|J_{xy}|\leq n\|J_{\mathrm{abs}}\|_{\mathrm{op}}<nN$.
\end{proof}

\begin{remark}[Entropic stability and $O(n\log n)$ mixing]\label{rem:nlogn}
Let $J_{\mathrm{abs}}=(|J_{xy}|)_{x,y}$ and
$\theta=\norm{J_{\mathrm{abs}}}_{\op}/N<1$.
Lemma~\ref{lem:two-spin} holds for arbitrary external fields and
couplings of either sign. Consequently, we can extend the previous proof to show that for every $s \in [0,1]$ and $\mathbf h \in \R^{nN}$, we have
\[
 \Cov_{\mu_{sJ,\mathbf h}}(\sigma)
 \preceq \frac{1}{N(1-s\theta)}I_{nN}.
\]
Having this, we may then apply the entropic stability framework of \cite{CE25} to get a modified log-Sobolev constant and $O(n\log n)$ mixing, essentially as obtained in \cite{AKV24}. We omit the details here because we think it will detract from the main point of this article to illustrate the extended Bochner method.  
\end{remark}

\subsection{Proof of the local inequality}\label{subsec:proof-local}
For $h\in\R^N$, we define the tilted Haar measure
$\nu_h(du)=Z_h^{-1}e^{h\cdot u}\,d\omega(u)$ on $\bS^{N-1}$, where
$Z_h=\int_{\bS^{N-1}}e^{h\cdot u}\,d\omega(u)$.
The Lipschitz constants below refer to Euclidean distance on the sphere. The key lemma that we need is the following.

\begin{lemma}\label{lem:covariance-bound}
For every $h,c\in\R^N$ and every Lipschitz function
$f:\bS^{N-1}\to\R$,
\begin{equation}\label{eq:covariance-bound}
\left|\Cov_{\nu_h}(f(U),c\cdot U)\right|
\le\frac{\norm c}{N}\Lip(f).
\end{equation}
\end{lemma}

Assuming Lemma~\ref{lem:covariance-bound}, we prove Lemma~\ref{lem:two-spin}.
\begin{proof}[Proof of Lemma~\ref{lem:two-spin}]
Let $\nu_U,\nu_V$ be the corresponding marginal laws, and define
$T:L^2(\nu_V)\to L^2(\nu_U)$ by
$(Tg)(u)=\E_\nu[g(V)\mid U=u]$.
Its adjoint is $(T^*f)(v)=\E_\nu[f(U)\mid V=v]$.
Both operators preserve constants and centeredness and as a consequence of Jensen, we also have $\norm T_{\op}\le1$. It remains to prove $\rho_\nu(U,V)\le|\kappa|/N$.

We write 
\begin{align*}
\frac{d}{ds}\E_{\nu_{h+sc}}g(U)
&=\frac{d}{ds}
  \left(
  \frac{\int_{\bS^{N-1}}g(u)e^{(h+sc)\cdot u}\,d\omega(u)}
       {Z_{h+sc}}
  \right)\\
&=\frac{\int_{\bS^{N-1}}g(u)(c\cdot u)e^{(h+sc)\cdot u}\,d\omega(u)}
        {Z_{h+sc}}\\
&\quad-
  \frac{\int_{\bS^{N-1}}g(u)e^{(h+sc)\cdot u}\,d\omega(u)}
       {Z_{h+sc}}
  \frac{\int_{\bS^{N-1}}(c\cdot u)e^{(h+sc)\cdot u}\,d\omega(u)}
       {Z_{h+sc}}\\
&=\E_{\nu_{h+sc}}[g(U)(c\cdot U)]
  -\E_{\nu_{h+sc}}[g(U)]\,
   \E_{\nu_{h+sc}}[c\cdot U]\\
&=\Cov_{\nu_{h+sc}}(g(U),c\cdot U).
\end{align*}
Since $(Tg)(u)=\E_{\nu_{b+\kappa u}}g(V)$, we integrate 
Lemma~\ref{lem:covariance-bound} along the segment
between $b+\kappa u$ and $b+\kappa u'$ to obtain
\begin{equation}\label{eq:Lip-contraction}
\Lip(Tg)\le\frac{|\kappa|}{N}\Lip(g),
\qquad
\Lip(T^*f)\le\frac{|\kappa|}{N}\Lip(f).
\end{equation}

We can write \[
(Tg)(u)=\int k(u,v)g(v)\,d\nu_V(v).
\]
where $k(u,v)=\frac{p(u,v)}{p_U(u)p_V(v)}
=\frac{d\nu}{d(\nu_U\otimes\nu_V)}(u,v)$, which is smooth and bounded since the density of $\nu$ is smooth and strictly positive on the compact
product of spheres. In particular, $T$ is
compact and $T,T^*$ map $L^2$ functions to smooth functions.
For $\rho:=\rho_\nu(U,V)>0$, compactness gives centered unit
singular functions $f,g$ satisfying $Tg=\rho f$ and
$T^*f=\rho g$. These functions are smooth and nonconstant, so their
Lipschitz constants are finite and positive. As a result, we can multiply 
\eqref{eq:Lip-contraction} to get
\[
\rho^2\Lip(f)\Lip(g)
\le\frac{\kappa^2}{N^2}\Lip(f)\Lip(g).
\]
Consequently, $\rho\le|\kappa|/N$ which completes the proof.
\end{proof}

Finally, we prove the estimate in Lemma~\ref{lem:covariance-bound}.

\begin{proof}[Proof of Lemma~\ref{lem:covariance-bound}]
Without loss of generality by scaling it suffices to consider $\Lip(f)\le1$.
Let $\mu$ be uniform probability measure on $\mathbb S^{N-1}$, and
write
\[
g(x)=\frac{e^{\langle h,x\rangle}}{Z_h}
\left(\langle c,x\rangle
-\mathbb E_{\nu_h}\langle c,X\rangle\right)
\quad\text{so that}\quad
\operatorname{Cov}_{\nu_h}(f(X),\langle c,X\rangle)
=\int_{\mathbb S^{N-1}}f(x)g(x)\,\mu(dx).
\]
We seek a vector field $V$ on $\mathbb B^N$ satisfying
\[
\operatorname{div}V=0
\quad\text{in }\mathbb B^N,
\qquad
V(x)\cdot x=g(x)
\quad\text{on }\mathbb S^{N-1}.
\]
With this vector field in hand, we can then ``transfer'' the derivative
onto $f$ via integration by parts, which would then allow us to exploit
the bound on its Lipschitz constant. Note here that by
Proposition~\ref{fact:divergence-lipschitz}, we have the necessary condition
of $\int_{\mathbb S^{N-1}}g(x)\,\mu(dx)=0$ for the divergence-free field.
We now carry out such a strategy.

We begin by writing $h=te$, where $t=\norm h$ and $\norm e=1$; when $t=0$,
choose any unit vector $e$. We have the decomposition $c=\alpha e+v$ with
$v\perp e$, and define
\[
\mathcal Z(s)=\int_{\bS^{N-1}}e^{s e\cdot u}\,d\omega(u),
\qquad
m(s)=\frac{\mathcal Z'(s)}{\mathcal Z(s)}.
\]
We observe that rotational invariance of the sphere gives
$Z_h=\mathcal Z(t)$ and $\E_{\nu_{se}}U=m(s)e$.
Moreover, the radial function
$y\mapsto\int e^{y\cdot u}\,d\omega(u)=\mathcal Z(\norm y)$
satisfies $\Delta_y\mathcal Z(\norm y)=\mathcal Z(\norm y)$.
Consequently, we have for $s>0$ that
\begin{equation}\label{eq:Z-ode}
\mathcal Z''(s)+\frac{N-1}{s}\mathcal Z'(s)=\mathcal Z(s)
 \quad\text{and}\quad
\mathcal Z(0)=1,\quad
\mathcal Z'(0)=0,\quad
\mathcal Z''(0)=\frac1N.
\end{equation}

For $0\le r\le1$, we define
\[
a(r)=\frac{\mathcal Z(tr)}{\mathcal Z(t)},\qquad
b(r)=\frac{\mathcal Z'(tr)}{r\mathcal Z(t)} \quad\text{and} \quad
b(0)=\frac{t}{N\mathcal Z(t)}.
\]
These functions extend smoothly as radial functions at the origin,
since $\mathcal Z$ is even and analytic. By \eqref{eq:Z-ode}, we have
$a'(r)=trb(r)$ and $rb'(r)+Nb(r)=ta(r)$.
Next, we define on the closed unit ball $\mathbb B^N$,
\[
V(z)=\frac{e^{h\cdot z}}{\mathcal Z(t)}
\left[v+\alpha\bigl(a(\norm z)e-b(\norm z)z\bigr)\right].
\]
For $r=\norm z>0$, we have
\begin{align*}
\diver V(z)
&=\frac{\alpha e^{h\cdot z}}{\mathcal Z(t)}
\left[
ta(r)-rb'(r)-Nb(r)
+(e\cdot z)\left(\frac{a'(r)}r-tb(r)\right)
\right]
=0.
\end{align*}
We can extend this to the origin by smoothness.
Since $a(1)=1$ and $b(1)=m(t)$, we have
\begin{equation}\label{eq:normal-trace}
V(u)\cdot u
=\frac{e^{h\cdot u}}{Z_h}
\left(c\cdot u-\E_{\nu_h}[c\cdot U]\right)
\qquad(u\in\bS^{N-1}).
\end{equation}

We next bound the integral of $\norm V$ over the ball. Fix $r\in[0,1]$
and put $s=tr$. Since $rb(r)=a(r)m(s)$,
\begin{align*}
\int_{\bS^{N-1}}\norm{V(ru)}\,d\omega(u)
&=a(r)\E_{\nu_{se}}
  \norm{v+\alpha a(r)(e-m(s)U)}\\
&\underset{
  \substack{
    \text{Cauchy-Schwarz}\\
    \E_{\nu_{se}}U=m(s)e,\,v\perp e
  }
}{\le}
a(r)\left(
  \norm v^2+\alpha^2a(r)^2(1-m(s)^2)
\right)^{1/2}\\
&\underset{|m(s)|\le1,\,0<a(r)\le1}{\le}\norm c,
\end{align*}
where $0<a(r)\le1$ follows because
$\mathcal Z(s)=\int\cosh(s e\cdot u)\,d\omega(u)$ is
nondecreasing for $s\ge0$. As a result,
\begin{align}\label{eq:vector-L1}
\int_{\mathbb B^N}\norm{V(z)}\,dz
=|\mathbb S^{N-1}|
  \int_0^1 r^{N-1}
  \left(
    \int_{\mathbb S^{N-1}}\norm{V(ru)}\,d\omega(u)
  \right)dr \le|\mathbb S^{N-1}|\norm{c}
  \int_0^1 r^{N-1}\,dr
  =|\mathbb B^N|\norm{c}.
\end{align}

Let $\widetilde f$ be a $1$-Lipschitz extension of $f$ to
$\mathbb R^N$, as provided by Proposition~\ref{prop:whitney-extension}. Then we write
\[\left|\Cov_{\nu_h}(f(U),c\cdot U)\right|
\underset{\eqref{eq:normal-trace}}{=}
\left|
\int_{\bS^{N-1}}\widetilde f(u)V(u)\cdot u\,d\omega(u)
\right| \underset{
  \text{Proposition~\ref{fact:divergence-lipschitz}, \eqref{eq:vector-L1}}
}{\leq}
\frac{|\mathbb B^N|}{N\omega_N}\norm{c}
=\frac{\norm{c}}{N}. \qedhere\]
\end{proof}

\section{Application: Negative correlation in matroids} \label{sec:neg-corr-full}
Let $M=(E,\mathcal I)$ be a finite matroid, and let
$\lambda\in(0,\infty)^E$ be a fugacity vector. We define a probability
measure on $\mathcal I$ by
\[
    \mu(S)=\frac{1}{Z}\prod_{e\in S}\lambda_e
    \quad\text{where}\quad
    Z=\sum_{S\in\mathcal I}\prod_{e\in S}\lambda_e.
\]
When $\lambda_e=1$ for every $e\in E$, this is the uniform
distribution on independent sets.

For $X\sim\mu$, we define $X_e=\mathbf 1_{\{e\in X\}}$,
$p_e=\E_\mu X_e$, and $p_{ef}=\E_\mu[X_eX_f]$.
Let $p=(p_e)_{e\in E}$ and
$\Sigma=(\Cov_\mu(X_e,X_f))_{e,f\in E}$.

We will first recover the following known covariance bound
for matroid independent sets using the abstract machinery relating to $B^{-1}$.
% Add the appropriate citation for this known result.
\begin{theorem}\label{thm:neg-corr-matroid}
    We have $\Sigma\preceq\operatorname{diag}(p)$.
    Consequently, for all distinct $e,f\in E$,
    \[
        p_{ef}\leq 2p_ep_f.
    \]
\end{theorem}

For graphic matroids, whose independent sets are forests, we obtain
the following stronger bound. As above, the fugacities may be
arbitrary positive numbers.
\begin{theorem}
    If $M$ is graphic, then for all distinct $e,f\in E$,
    \[
        p_{ef}\leq \frac{4}{3}p_ep_f.
    \]
\end{theorem}

\subsection{Notation} \label{subsec:neg-corr-notation}

We now adapt our Hilbert space set-up to the above measure on independent sets of a matroid. First, we let $\mathcal H^0=L^2(\mu)$ and define
\[
    W_e=\operatorname{ran}L_e,\qquad
    W_{ef}=W_e\cap W_f,\qquad
    \mathcal H^1=\bigoplus_eW_e \quad\text{and}\quad
    \mathcal H^2=\bigoplus_{e<f}W_{ef}.
\]
As before, let $W=\sum_eW_e$ and let $P$ be the orthogonal projection
onto $W$.

For this application, we define the weights $c_e=1+\lambda_e$ and corresponding
weighted inner products
\begin{equation}\label{eq:matroid-inner-products}
    \langle u,v\rangle_1
       =\sum_e c_e\langle u_e,v_e\rangle_0,
    \quad\text{and}\quad
    \langle a,b\rangle_2
       =\sum_{e<f}c_ec_f\langle a_{ef},b_{ef}\rangle_0,
\end{equation}
where we recall that $\langle h,g\rangle_0=\E_\mu[hg]$ as before. 

As before we have $d_0:\mathcal H^0\to\mathcal H^1$ and
$d_1:\mathcal H^1\to\mathcal H^2$, given by $(d_0h)_e=L_eh$ and  $ (d_1u)_{ef}=L_{ef}(u_e-u_f)$,
where $L_{ef}$ is the orthogonal projection onto $W_{ef}$ and $L_e = I - P_e$. Similarly as before, we have a chain complex $d_1d_0=0$. With the weighted inner products, we have
\[
    d_0^*u=\sum_e c_e u_e,
    \qquad
    Q=d_0^*d_0=\sum_e c_eL_e.
\]

For $S\in\mathcal I$, we also define
$A(S)=\{e\notin S:S\cup\{e\}\in\mathcal I\}$ to be the ``available'' set of elements to add to $S$. We write
\begin{equation}\label{eq:matroid-laplacian}
    (Qh)(S)
       =\sum_{e\in S}\bigl(h(S)-h(S-e)\bigr)
        +\sum_{e\in A(S)}\lambda_e
           \bigl(h(S)-h(S\cup\{e\})\bigr)
\end{equation}
Since $\mu(S)\lambda_e=\mu(S\cup\{e\})$ whenever $e\in A(S)$,
it follows that
\begin{equation}\label{eq:matroid-dirichlet}
    \langle h,Qg\rangle_0 =\langle d_0h,d_0g\rangle_1 =\sum_{S\in\mathcal I}\mu(S)
          \sum_{e\in A(S)}\lambda_e
          \bigl(h(S\cup\{e\})-h(S)\bigr)
          \bigl(g(S\cup\{e\})-g(S)\bigr).
\end{equation}

Since all fugacities are positive,
\eqref{eq:matroid-dirichlet} shows that $\ker Q$ consists
only of constant functions. Having this, we get $W^\perp=\bigcap_e\ker L_e=\ker Q$ and $W=L^2_0(\mu)$, so that in particular $Ph = h - \E_{\mu} h$. 

\paragraph{Bochner operator.}
As in Lemma~\ref{lem:Bochner-lite}, we define
$B:\mathcal H^1\to\mathcal H^1$ by $ B=d_0d_0^*+d_1^*d_1$. Explicitly, we can write $(Bu)_e
       =L_e\left(\sum_f c_f u_f\right)
        +\sum_{f\ne e}c_fL_{ef}(u_e-u_f)$, i.e.  the blocks $B_{ef}:W_f\to W_e$ are
\begin{equation}\label{eq:matroid-bochner-blocks}
    B_{ef}=
    \begin{cases}
    c_eI+\displaystyle\sum_{k\ne e}c_kL_{ek},&e=f,\\[4pt]
        c_f(L_eL_f-L_{ef}),&e\ne f.
    \end{cases}
\end{equation}

Indeed,
whenever $B$ is invertible, the identity $P=d_0^*B^{-1}d_0$ from
Lemma~\ref{lem:Bochner-lite} gives
\begin{equation}\label{eq:matroid-inverse-covariance}
    \Cov_\mu(h,g)
       \underset{Ph=h-\E_\mu h}{=} \langle Ph,Pg\rangle_0
       =\langle d_0h,B^{-1}d_0g\rangle_1.
\end{equation}

\subsection{Matroids: Proof of Theorem~\ref{thm:neg-corr-matroid}}
\label{sec:matroid-correlation}

In this section, we expand on the intuition that the off-diagonal
blocks $B_{ef}=c_f(L_eL_f-L_{ef})$ capture interactions between
coordinates that control negative correlations. To that end,
we prove the bound $p_{ef}\leq2p_ep_f$ for general matroids,
illustrating how correlation bounds arise naturally from
estimates on the Bochner operator.

\begin{proposition}\label{prop:matroid-bochner-inverse}
    We have $B\succeq I$. Consequently, $B$ is invertible
    and $B^{-1}\preceq I$.
\end{proposition}

Assuming Proposition~\ref{prop:matroid-bochner-inverse}, we prove
Theorem~\ref{thm:neg-corr-matroid}.

\begin{proof}[Proof of Theorem~\ref{thm:neg-corr-matroid}]
Fix $a\in\mathbb R^E$ and let $h(S)=\sum_{e\in S}a_e$.
By Proposition~\ref{prop:matroid-bochner-inverse}, we may apply
Lemma~\ref{lem:Bochner-lite} with $\alpha=1$. We write
\begin{align*}
    a^\top\Sigma a
    =\Var_\mu(h) \underset{\text{Lemma~\ref{lem:Bochner-lite}}}{\leq}
      \langle h,Qh\rangle_0\underset{\eqref{eq:matroid-dirichlet}}{=}
      \sum_{S\in\mathcal I}\mu(S)
      \sum_{e\in A(S)}\lambda_e a_e^2 \underset{\mu(S)\lambda_e=\mu(S\cup\{e\})}{=}
      \sum_e a_e^2
      \sum_{\substack{T\in\mathcal I\\e\in T}}\mu(T)
      =\sum_e p_e a_e^2.
\end{align*}
Consequently, we have $\Sigma\preceq\operatorname{diag}(p)$.

Define $\Delta=\operatorname{diag}(p)-\Sigma\succeq0$.
Since $\Delta_{ee}=p_e-\Var_\mu(X_e)=p_e^2$, we have
\[
    p_{ef}
    =p_ep_f-\Delta_{ef}
    \underset{\text{$2\times2$ principal minors}}{\leq}
    p_ep_f+\sqrt{\Delta_{ee}\Delta_{ff}}
    =2p_ep_f.
\]
This completes the proof.
\end{proof}

Finally, we prove Proposition~\ref{prop:matroid-bochner-inverse}.

\begin{proof}[Proof of Proposition~\ref{prop:matroid-bochner-inverse}]
For each $S\in\mathcal I$, let $\mathcal P(S)$ be the partition
of $A(S)$ into parallel classes of $M/S$.

For $u_e\in W_e$, we have
$u_e(S)=-\lambda_e u_e(S\cup\{e\})$ whenever $e\in A(S)$,
and $u_e(S)=0$ when $e\notin S\cup A(S)$. Consequently,
\begin{equation}\label{eq:matroid-component-norm}
    \sum_{\substack{S\in\mathcal I\\e\in A(S)}}
    \mu(S)u_e(S)^2
    =\frac{\lambda_e}{c_e}\|u_e\|_0^2.
\end{equation}

Next, we condition on all coordinates except $e,f$. 
If $S\cup\{e,f\}$ is independent, then the conditional law is a product measure and $L_eL_f=L_{ef}$. If either element cannot be added to $S$,
the corresponding projection is zero. 

So it remains to account for the case where we can add either $\{e \}$ or $\{f\}$ which corresponds to when $e,f$ belong to the same parallel class of $M/S$.  In this case, we write
\begin{align}\label{eq:Lij}
    (L_{ef}u_f)(S) &=-\lambda_e(L_{ef}u_f)(S\cup\{e\})\underset{L_fL_{ef}=L_{ef}}{=}-\lambda_e\bigl[(I-P_f)L_{ef}u_f\bigr](S\cup\{e\})\\&=-\lambda_e\bigl(
        (L_{ef}u_f)(S\cup\{e\})
        -(L_{ef}u_f)(S\cup\{e\})
      \bigr)
    =0.
\end{align}

It follows that the term in $\langle u, (B-I) u\rangle_1$ coming from $S, S \cup \{e \}$ and $ S\cup \{f\}$ is
\begin{align*}
    c_e\sum_{T\in\{S,S\cup\{e\},S\cup\{f\}\}}
      \mu(T)u_e(T)(B_{ef}u_f)(T) &\underset{
        \eqref{eq:matroid-bochner-blocks}, \, \eqref{eq:Lij},\,
        L_eu_e=u_e,\;L_fu_f=u_f,
      }{=}
      c_ec_f\sum_{T\in\{S,S\cup\{e\},S\cup\{f\}\}}
      \mu(T)u_e(T)u_f(T)\\
    &\underset{
        u_e(S\cup\{f\})=u_f(S\cup\{e\})=0
      }{=}
      c_ec_f\mu(S)u_e(S)u_f(S).
\end{align*}

Consequently, summing up the above over all $S$ and $e,f$, we have for $u=(u_e)_e\in\mathcal H^1$ that 
\begin{align}\label{eq:matroid-main}
    \langle u,(B-I)u\rangle_1
    &\underset{
        \eqref{eq:matroid-bochner-blocks}
      }{\geq}
      \sum_e c_e\lambda_e\|u_e\|_0^2 
      +\sum_{S\in\mathcal I}\mu(S)
       \sum_{C\in\mathcal P(S)}
       \sum_{\substack{e,f\in C\\e\ne f}}
       c_ec_f u_e(S)u_f(S)\notag\\
    &\underset{\eqref{eq:matroid-component-norm}}{=}
      \sum_{S\in\mathcal I}\mu(S)
      \sum_{C\in\mathcal P(S)}
      \left(\sum_{e\in C}c_e u_e(S)\right)^2
      \geq0.
\end{align}

This shows that $B\succeq I$. Lemma~\ref{lem:Bochner-lite} supplies the
invertibility of $B$, and consequently we also have $B^{-1}\preceq I$.
\end{proof}

\subsection{Forests: Proof of Theorem~\ref{thm:neg-corr-forest}} \label{subsec:forests}
We now turn from the general matroid set-up to the setting of forests, where the additional structure in graphic matroids allows us to sharpen the correlation bound. For spanning trees, we recall that negative correlation can be proven via the determinantal representation and real stability of the spanning-tree polynomial. As described in the technical overview, our strategy is to extend this determinantal representation to the setting of forests. 

While showing log-concavity of this auxiliary polynomial is the driving force of this proof, one way to compare this with the analysis of Section~\ref{sec:matroid-correlation} is that here this real stability provides a more fine-grained bound of quadratic forms of $B^{-1}$ acting on $\mathrm{im}(d_0)$. 

In the following, we define 
\[
    s_{ef}
    =\Pr_\mu\bigl(e,f\text{ are edges of the same
       tree component of }F\bigr)\quad\text{and}\quad
     S=(s_{ef})_{e,f\in E}.
\]
Particularly, 
$s_{ee}=p_e$ and $s_{ef}\leq p_{ef}$.

\begin{theorem}\label{thm:neg-corr-forest}
We have
\begin{equation}\label{eq:forest-covariance-matrix}
    \Sigma\preceq\frac34\diag(p)+\frac14 S.
\end{equation}
Moreover, for all distinct $e,f\in E$,
\begin{equation}\label{eq:forest-covariance-signed}
    4(p_{ef}-p_ep_f)\le s_{ef}\le p_{ef}.
\end{equation}
Consequently,
\begin{equation}\label{eq:forest-four-thirds}
    p_{ef}\le\frac43p_ep_f.
\end{equation}
\end{theorem}

The following result gives the bounds on
$\langle d_0X_e,B^{-1}d_0X_f\rangle$ (rather than on all of $\mathcal{H}^1$)
needed below.

\begin{proposition}\label{prop:forest-bochner-inverse}
For every $\mathbf{a}\in\mathbb R^E$ and $h=\sum_e a_eX_e$, we have
\begin{equation}\label{eq:forest-inverse-quadratic}
    4\langle d_0h,B^{-1}d_0h\rangle_1
    \le \mathbf{a}^\top\bigl(3\diag(p)+ S\bigr)\mathbf{a}.
\end{equation}
Moreover, for all distinct $e,f\in E$,
\begin{equation}\label{eq:forest-inverse-bound}
    4c_e\langle L_eX_e,(B^{-1})_{ef}L_fX_f\rangle_0
    \le s_{ef}.
\end{equation}
\end{proposition}

Assuming Proposition~\ref{prop:forest-bochner-inverse}, we prove
Theorem~\ref{thm:neg-corr-forest}.

\begin{proof}[Proof of Theorem~\ref{thm:neg-corr-forest}]

Fix $\mathbf{a}\in\mathbb R^E$ and let $h=\sum_{e\in E}a_eX_e$.
Then
\[
    \mathbf{a}^\top\Sigma \mathbf{a}
      =\langle d_0h,B^{-1}d_0h\rangle_1
      \le\frac14
         \mathbf{a}^\top\bigl(3\operatorname{diag}(p)+S\bigr)\mathbf{a}.
\]
Since the choice of $a$ is arbitrary, it follows that \eqref{eq:forest-covariance-matrix} holds.

For distinct $e,f$, the only nonzero coordinate of $d_0X_e$
is its $e$th coordinate, so
\begin{align*}
    4(p_{ef}-p_ep_f) \underset{\eqref{eq:matroid-inverse-covariance}}{=}
      4c_e\langle L_eX_e,(B^{-1})_{ef}L_fX_f\rangle_0 \underset{\eqref{eq:forest-inverse-bound}}{\le}s_{ef}
      \le p_{ef}.
\end{align*}
Rearranging gives the desired inequality $p_{ef}\le\frac43p_ep_f$.
\end{proof}

The main step is to study a polynomial that in some sense generalizes the determinant construction used to prove negative correlation
for spanning trees; we define
\begin{equation}\label{eq:component-matrix}
     A_F
    =\diag(X)+\sum_{T\in\Comp(F)}
      \one_{E(T)}\one_{E(T)}^\top \quad\text{and}\quad
    \Phi(z)=\E_\mu\det\!\bigl(I+\diag(z) A_F\bigr).
\end{equation}

\begin{proposition}\label{prop:component-stability}
The polynomial $\Phi$ is real stable.
\end{proposition}

Assuming Proposition~\ref{prop:component-stability}, we prove Proposition~\ref{prop:forest-bochner-inverse}. 

\begin{proof}[Proof of Proposition~\ref{prop:forest-bochner-inverse}]

We claim that
\begin{equation}\label{eq:forest-inverse-response}
    \nabla^2\log\Phi(0)
    =4\Bigl(c_e
        \langle L_eX_e,(B^{-1})_{ef}L_fX_f\rangle_0
      \Bigr)_{e,f\in E}
      -3\diag(p)- S,
\end{equation}
the proof of which we defer to the end. For now we use this identity and the real stability of $\Phi$ as in Proposition~\ref{prop:component-stability} to deduce the bounds on $B^{-1}$.

Since $A_F\succeq0$, the principal-minor expansion shows that
$\Phi$ is multiaffine with nonnegative coefficients and $\Phi(0)=1$. Proposition~\ref{prop:component-stability} consequently implies that for $e\neq f$ that \begin{equation}\label{eq:component-hessian-signs}
    \nabla^2\log\Phi(0) \underset{Fact~\ref{fact:stable-log-concave}}{\preceq} 0 \quad\text{and}\quad
    \partial_e\partial_f\log\Phi(0)\underset{Fact~\ref{fact:rayleigh}}{\le} 0.
\end{equation}
We let $\mathbf{a}\in\mathbb R^E$ and $h=\sum_e a_eX_e$. Since only the $e$th coordinate of $d_0X_e$
is nonzero, we have
\begin{align*}
    4\langle d_0h,B^{-1}d_0h\rangle_1 \underset{\eqref{eq:forest-inverse-response}}{=}
      \mathbf{a}^\top\bigl(
        \nabla^2\log\Phi(0)+3\diag(p)+ S
      \bigr) \mathbf{a} \underset{\eqref{eq:component-hessian-signs}}{\leq}
      \mathbf{a}^\top\bigl(3\diag(p)+ S\bigr)\mathbf{a}.
\end{align*}
This proves \eqref{eq:forest-inverse-quadratic}. For distinct
$e,f$, we similarly obtain
\begin{align*}
    4c_e\langle L_eX_e,(B^{-1})_{ef}L_fX_f\rangle_0 \underset{\eqref{eq:forest-inverse-response}}{=}
      \partial_e\partial_f\log\Phi(0)+s_{ef} \underset{\eqref{eq:component-hessian-signs}}{\leq}s_{ef},
\end{align*}
which proves \eqref{eq:forest-inverse-bound}.

It remains to prove \eqref{eq:forest-inverse-response}. For $z$
near the origin, define
\[
    H(z,F)=\log\det\!\bigl(I+\diag(z) A_F\bigr),
    \qquad
    \Phi(z)=\E_\mu e^{H(z,F)}.
\]
Since $H(0,F)=0$, differentiating the finite expectation gives
\[
    \nabla^2\log\Phi(0)
    =\E_\mu\nabla^2H(0,F)
      +\Cov_\mu\bigl(\nabla H(0,F)\bigr).
\]
Since $\partial_eH(0,F)=( A_F)_{ee}=2X_e$ and $\partial_e\partial_fH(0,F)
    =-( A_F)_{ef}^2$, we have
\[
    \E_\mu\partial_e\partial_fH(0,F)
    =
    \begin{cases}
        -4p_e,&e=f,\\
        -s_{ef},&e\ne f.
    \end{cases}
\]
Since $s_{ee}=p_e$, we consequently have $\E_\mu\nabla^2H(0,F)=-3\diag(p)- S$. It follows that $\nabla^2\log\Phi(0)
    =-3\diag(p)- S+4\Cov_\mu(X)$ which rearranges to \eqref{eq:forest-inverse-response} after \eqref{eq:matroid-inverse-covariance} to each
entry of $\Cov_\mu(X)$.
\end{proof}

\begin{proof}[Proof of Proposition~\ref{prop:component-stability}]
For an edge $e=uv$, we write $b_e=\mathbf e_u-\mathbf e_v$ and
$\mathbf{x}=(x_v)_{v\in V}$. We also define
\[
 R(\mathbf{x},z)=
 \det\!\left(
     \operatorname{diag}(\mathbf{x})
     +\sum_{e\in E}\lambda_e(1+z_e)b_eb_e^\top
 \right),
 \qquad
 \mathcal T=
 \prod_{e=uv\in E}
 \left(1-\lambda_e\partial_{x_u}\partial_{x_v}\right).
\]
The key identity is
\begin{equation}\label{eq:component-stability-operator}
    Z\Phi(z)=\left.\mathcal T R(\mathbf{x},z)\right|_{\mathbf{x}=\mathbf1}.
\end{equation}
Assuming this identity, we prove the real stability of $\Phi$.
First, $R$ is jointly real stable in $\mathbf{x},z$: when all variables
have positive imaginary parts, the matrix defining $R$ has
positive definite imaginary part
\[
 \operatorname{diag}(\operatorname{Im}\mathbf{x})
 +\sum_{e\in E}\lambda_e\operatorname{Im}(z_e)b_eb_e^\top
 \succ0,
\]
and is therefore nonsingular.

Moreover, $R$ is multiaffine in $\mathbf{x},z$, so each factor of
$\mathcal T$ preserves stability by
Fact~\ref{fact:stable-pair-derivative}.
The resulting polynomial is still multiaffine, and its Rayleigh
inequalities remain valid after setting $\mathbf{x}=\mathbf1$.
Fact~\ref{fact:rayleigh} and
\eqref{eq:component-stability-operator} therefore show that
$\Phi$ is real stable.

Finally, we prove \eqref{eq:component-stability-operator}.
To that end, we first apply the rooted-forest form of the
weighted matrix-tree theorem to write
\[
 R(\mathbf{x},z)=
 \sum_{H\in\mathcal F(G)}
 \prod_{e\in H}\lambda_e(1+z_e)
 \prod_{C\in\operatorname{Comp}(H)}
       \left(\sum_{v\in V(C)}x_v\right).
\]
When we expand the final product, we correspondingly choose a root for
each connected component of $H$.

In the expansion of $\mathcal T$, selecting the term
$-\lambda_e\partial_{x_u}\partial_{x_v}$ requires $u$ and $v$
to be the roots of different connected components.
We interpret its action as erasing these variables corresponding
to the roots and instead joining the two components by
$e=uv$ with weight $-\lambda_e$.
Since each component has only one root, it can participate in at
most one such joining.

After setting $\mathbf{x}=\mathbf1$, a tree $T$ is formed in the
following two ways:
\begin{enumerate}
    \item (We did not add any edge to join two components.) The original component was
    $T$, and its root could be any of its $|V(T)|$ vertices.
    These choices contribute
    $|V(T)|\prod_{f\in E(T)}\lambda_f(1+z_f)$.

    \item (We added an edge $e=uv$ to join two components.)
    We recover the two disjoint connected components as the two
    trees obtained by deleting $e$ from $T$, rooted at $u$ and $v$.
    For each $e\in E(T)$, this determines one such unique contribution
    of
    \[
        -\lambda_e
        \prod_{f\in E(T)\setminus\{e\}}\lambda_f(1+z_f).
    \]
\end{enumerate}
Summing these possibilities, the contribution of $T$ is
\begin{equation}\label{eq:component-product}
    \lambda^{E(T)}
 \left[
   |V(T)|\prod_{e\in E(T)}(1+z_e)
   -
   \sum_{e\in E(T)}
      \prod_{f\in E(T)\setminus\{e\}}(1+z_f)
 \right].
\end{equation}
The choices in distinct connected components are independent,
so their contributions multiply.
For $S\subseteq E(T)$, the coefficient of $z^S$ in the expression above is $|V(T)|-|E(T)\setminus S|=1+|S|$.

On the other hand, the block of $ A_F$ indexed by $E(T)$
is $I+\mathbf1\mathbf1^\top$, whose principal minor on $S$ is $\det(I_{|S|}+\mathbf1\mathbf1^\top)=1+|S|$. 

From a principal-minor expansion,
for every $S\subseteq F$ the coefficient of $z^S$ in
$\det(I+\operatorname{diag}(z)A_F)$ is
$\prod_{T\in\operatorname{Comp}(F)}
        \bigl(1+|S\cap E(T)|\bigr)$
due to the block structure of $A_F$.
As we have seen, this is also the coefficient of $z^S$ in
\eqref{eq:component-product}. Taking a product of all connected components and summing over forests completes the proof of
\eqref{eq:component-stability-operator}.
\end{proof}

\section{AI Declaration}

We used OpenAI’s ChatGPT extensively in the mathematical development and preparation of this paper. Our use included proposing and refining mathematical formulations, exploring proof strategies, developing and checking candidate arguments, working through calculations and examples, identifying relevant literature, and drafting and revising exposition and notation. Some of these exchanges produced substantial derivations and proposed complete proofs that influenced the statements and arguments presented in the paper. The use of ChatGPT was iterative and interwoven with the authors’ research and writing, as described below.

\paragraph{Hilbert-space framework and general results.}

ChatGPT assisted in developing the general formulation in terms of closed subspaces and orthogonal projections on a possibly infinite-dimensional Hilbert space, and in extending the relevant finite-spin results to this setting. This included developing and explaining the maps $d_0$ and $d_1$, the Bochner operator $B=d_0d_0^*+d_1^*d_1$, its relationship to the Laplacian and the Bochner and Poincar\'e inequalities, and the covariance representation involving $B^{-1}$. We also used ChatGPT to work through the functional-analytic details and hypotheses, including orthogonal complements and intersections, closures of subspace sums and operator ranges, bounded invertibility, and the connectedness assumptions needed in the spin-system applications. Related discussions addressed conditional expectations, maximal correlation, and the relationships and normalization differences between Glauber dynamics and the associated bipartite operators. These exchanges materially influenced the formulations, proofs, and explanations in Sections \ref{sec:preliminaries} and \ref{sec:orthog-proj}.

The assistance also extended to the formulation and proof of the general block result. Following our questions about passing from individual coordinates to disjoint groups of coordinates, ChatGPT proposed and developed restriction and compression arguments, Schur-complement comparisons, and estimates for the resulting operator blocks. It supplied substantial candidate proof material for the general projection theorem and its application to the spectral gaps of marginal distributions grouped into blocks. We examined and revised this material as the generalization was incorporated into the main results of the paper. ChatGPT also helped develop the Gaussian interpretation of the comparison matrix and the examples showing sharpness of the bounds. The research questions and earlier results supplied by the authors guided these exchanges; the resulting assistance affected both the statements and the proofs of the generalization.

\paragraph{Negative correlations in matroids and forests.}

ChatGPT assisted substantially with the mathematical development of Section~\ref{sec:neg-corr-full}. This included calculations and proof arguments involving the weighted Bochner operator, covariance identities involving its inverse, and the recovery of the factor $2$ bound on pairwise inclusion probabilities for random matroid independent sets. In particular, in response to our request to generalize the determinantal proof of negative correlation for spanning trees using the Bochner framework, ChatGPT suggested the approach based on real stability that yields the $\frac{4}{3}$ correlation bound for forests. Subsequent exchanges developed and checked the auxiliary polynomial argument, the resulting covariance and pairwise correlation estimates, and their connection to the inverse Bochner operator. These arguments and their presentation were revised during the preparation of the paper.

\paragraph{The $O(N)$ application.}

In Section~\ref{sec:full-ON-app}, ChatGPT assisted in developing the radial ansatz and coefficient functions for the divergence-free vector field used to prove the covariance bound for tilted spherical measures. This is a substantive ingredient of the local inequality underlying the $O(N)$ application. We also used ChatGPT to review details and explanations of the argument and its connection to the two-site maximal-correlation estimate, the global spectral-gap bound, and the mixing-time consequences.

\paragraph{Related literature, interpretations, and further consequences.}

We used ChatGPT to identify and examine relevant literature and to compare the hypotheses, normalizations, and conclusions of related results. This included comparisons with existing Bochner and trickle-down arguments, covariance representations, and results on angles between subspaces, including those of Kassabov. ChatGPT also assisted in deriving and presenting consequences and comparisons involving $\lambda$-products, swap walks, path complexes, and random flags of subspaces, including the role of retaining the full matrix of conditional pairwise bounds when analyzing groups of coordinates. These discussions informed the applications in Section~\ref{sec:intro-hdx} and the explanations relating the different spectral and correlation quantities in Section~\ref{subsec:spectral-gaps}. Literature suggestions and comparisons were considered as part of the authors’ assessment of the relationship to prior work.

\paragraph{Writing, review, and author responsibility.}

We used ChatGPT to review successive manuscript drafts, suggest corrections and missing explanations, and help revise the abstract, introduction, technical overview, proofs, notation, and LaTeX. Earlier drafts also included AI-assisted transcriptions of author-supplied mathematical notes. The assistance included generating draft passages, proposing alternative organizations and explanations, and shortening or expanding arguments during revision. We also used the tool to explore alternative proof approaches and broader connections, including directions that were not retained in the paper.

Throughout this process, the authors supplied questions, existing results, draft statements, arguments, and manuscript versions; examined the responses; and developed, corrected, or discarded suggestions. Some AI-generated proposals contained errors, overlooked hypotheses, or did not lead to complete arguments. The authors selected and revised the material included in the paper and take full responsibility for its accuracy, originality, and integrity, including its mathematical statements, proofs, and references and all material developed with AI assistance. ChatGPT also assisted in preparing this declaration.

\bibliographystyle{alpha}
\bibliography{ref.bib}
\end{document}